\documentclass[aip,jcp,reprint]{revtex4-1}
\usepackage[english]{babel}
\usepackage{amsmath}
\usepackage{amsfonts}
\usepackage{graphicx}
\usepackage{amssymb}
\usepackage{soul}
\usepackage{ulem}
\usepackage{mdframed}
\usepackage{appendix}
\usepackage{hyperref}

\usepackage{color}

\newcommand*{\eq}[1]{Eq.\,\ref{eq:#1}}
\newcommand*{\eqtwo}[2]{Eqs.\,\ref{eq:#1},\ref{eq:#2}}
\newcommand*{\eeqq}[2]{\begin{equation} #1 \label{#2} \end{equation}}
\newcommand*{\eeqqa}[2]{\begin{align} \centering #1 \label{#2} \end{align}}

\newcommand*{\kbt}{\ensuremath{k_{\rm B} T}}

\newcommand*{\DGcnf}{\Delta G^{\rm cnf}}
\newcommand*{\DGcnfrR}{\Delta G^{\rm cnf}\left({\bf r}_i,{\bf r}_j,\{{\bf R}_i\}\right)}

\newcommand*{\etal}{{\it et al }}

\renewcommand{\vec}[1]{\mathbf{#1}}

\newcommand{\DG}{\Delta G}

\begin{document}

%\title{How Theory and Simulation can guide the rational design of DNA-coated colloids}
\title{Theory and Simulation of DNA-Coated Colloids: a Guide for Rational Design}
\author{Stefano Angioletti-Uberti}
%\email{stefano@buct.edu.cn}
\thanks{These authors contributed equally to this work}
\affiliation{International Research Centre for Soft Matter, Beijing University of Chemical Technology, 100029 Beijing, PR China}
\author{Bortolo M. Mognetti}
%\email{bmognett@ulb.ac.be}
\thanks{These authors contributed equally to this work}
\affiliation{Center for Nonlinear Phenomena and Complex Systems, Universit\'{e} Libre de Bruxelles, Code Postal 231, Campus Plaine, B-1050 Brussels, Belgium}
\author{Daan Frenkel}
\thanks{Corresponding author}
\email{df246@cam.ac.uk}
\affiliation{Department of Chemistry, University of Cambridge, Lensfield Road, CB2 1EW Cambridge, UK}

\begin{abstract}
By exploiting the exquisite selectivity of DNA hybridization, DNA-Coated Colloids (DNACCs) can be made to self-assemble in a wide variety of structures. 
The beauty of this system stems largely from its exceptional versatility and from the fact that a proper choice of the grafted DNA sequences yields fine control over the  colloidal interactions. 
Theory and  simulations have an important role to play in the optimal design of self-assembling DNACCs. At present, the powerful model-based design tools are not widely used, 
because the theoretical literature is fragmented and the connection between different theories is often not evident.
In this Perspective, we aim to discuss the similarities and differences between the different models that have been described in the literature, 
their underlying assumptions, their strengths and  their weaknesses. 
%By also discussing the link between such models and selected experimental results appeared in the literature, we will try to highlight 
%some general concepts related to DNA-mediated interactions whose full appreciation we believe will help the design of future systems, and 
%suggest possible new experiments. 
%
Using the tools described in the present Review, it should be possible to move towards a more rational design of novel  self-assembling structures 
 of DNACCs and, more generally, of systems  where ligand-receptors bonds are used to control interactions.
\end{abstract}

\maketitle

\section{Introduction \label{sec:Introduction}}
Colloidal suspensions are well described by the same statistical mechanical equations as systems of atoms or small molecules. 
As a consequence, colloids behave in many respects as `scaled-up' models of atoms, and like atoms, colloidal suspensions can be 
found in phases that resemble gases, liquids or crystals. There are, however, important differences. One is that colloids  move in 
a solvent, rather than in vacuum. The other is that, through modification of the colloid or the solvent, we can tune the interactions 
between colloids in a way that is not possible for atoms or small molecules. Due to our ability to control colloidal interactions, colloids 
exhibit a much richer phase behaviour than atomic systems. Over the past two decades, our ability to design colloidal interactions 
has undergone a quantum jump with the development of so-called DNA-Coated Colloids (DNACCs), colloidal particles functionalised 
with short sequences of single-stranded DNA (ssDNA). DNA allows the colloids to bind selectively, either directly or through single-stranded 
linkers,  to other particles or surfaces coated with  complementary ssDNA sequences. 

DNA-Coated Colloids were developed by the groups of Mirkin and Alivisatos \cite{mirkin,alivisatos} in the 1990s. Over the past decades, 
DNACCs   (sometimes referred to as  ``spherical nucleic acid'' \cite{mirkin-sna}) have been studied intensively because of their promise 
for the development of  new classes of ordered colloidal materials (see e.g.~\onlinecite{crystal-mirkin,crystal-gang,gang-switch,gang-multiple-particles}).
Furthermore, this system has interesting biomedical applications as ultra-sensitive biomarkers, detectors or efficient drug- and gene-delivery carriers 
\cite{taton,mirkin-gene1,mirkin-gene2}. In all cases, applications exploit the selectivity of DNA-DNA hybridization between a complementary pair, 
as well as specific features of the physics of multivalent interactions.\\
Several reviews on DNACCs have appeared in the literature, both from an experimental \cite{lorenzo,geerts-review} and from a 
theoretical \cite{dellago-review,travesset-review} perspective. Very recently, Jones \etal have provided a general overview describing the use of 
DNA-mediated interactions  for the programmable self-assembly of nanomaterials \cite{seeman-mirkin-review}, highlighting the connections, and 
the differences, between DNACCs  and DNA-based nanotechnology. 

The focus of the present Perspective is different. It is not our aim to provide a comprehensive overview of all simulation and modelling results in the field,  but rather, to present a critical assessment of the various theoretical and computational approaches that have been proposed to describe DNACCs. However, this Perspective is  not just a Review: to illustrate the points that we make, we also include new results.
We aim to highlight  the connections between different theoretical models, and explain the differences between various coarse-graining strategies 
used to describe DNACCs. 
We believe that the present perspective is timely, as the rational design of novel DNACC-based materials will require a complete integration  of theory and experiment.\\ 
At present,  theory and computational work are mostly used to explain experimental observations  {\it a posteriori}. Yet these experiments 
are still largely driven  by empirical or semi-empirical rules \cite{crystal-mirkin2}. Even when model predictions preceded experiments, the link between theory 
and experiments is often less than clear.\\
At present,  there is a bewildering variety of DNACC models in the literature, and this very fact makes it difficult for the non-expert to decide what approach 
to use. The net result is that most models are never used by experimentalists. 
More worrying is the fact that theoretical models that have been shown to be based on questionable premises continue to be used in the literature, thus carrying the risk that analysis of the experimental data is flawed.
For these reasons, the aim of this review is to discuss the various models in terms of their underlying  assumptions, their range of applicability 
and their relative strengths and weaknesses, so that experimentalists can use them in an informed way. 
\\
The remainder of the paper will be structured as follows. In Sec.~\ref{sec:theory}, we describe the general principles common 
to the various analytical theoretical models, and then proceed to discuss the connections between them. In particular, we aim to establish a 
`hierarchy of accuracy' in terms of the effects included. These analytical models contain input parameters that depend on the molecular structure 
of the system, e.g. the specific DNA-sequence grafted on the colloids, or details of the spacer used for grafting. When a systematic study of the 
effect of these parameters is required, and in order to relax some simplifying assumptions necessarily included in the analytical models, one should 
resort to computer experiments, i.e. simulations. We will deal with these ``experiments'' in Sec.~\ref{sec:simulations}, where we will focus on the 
various coarse-graining strategies that have been proposed in the literature. In particular, we discuss which (atomistic, molecular) features are retained 
in the various models, and which are ignored. In particular, we will focus on the question if the model can yield meaningful predictions of the 
relevant experiments. In the concluding section (Sec.~\ref{sec:conclusions}), we will  identify possible new directions where the existing theoretical 
and computational framework could be profitably integrated with experiments to yield a more rational design of DNACC-based systems.

\begin{figure}[h!]
\begin{center}
\includegraphics[width=1.1\columnwidth]{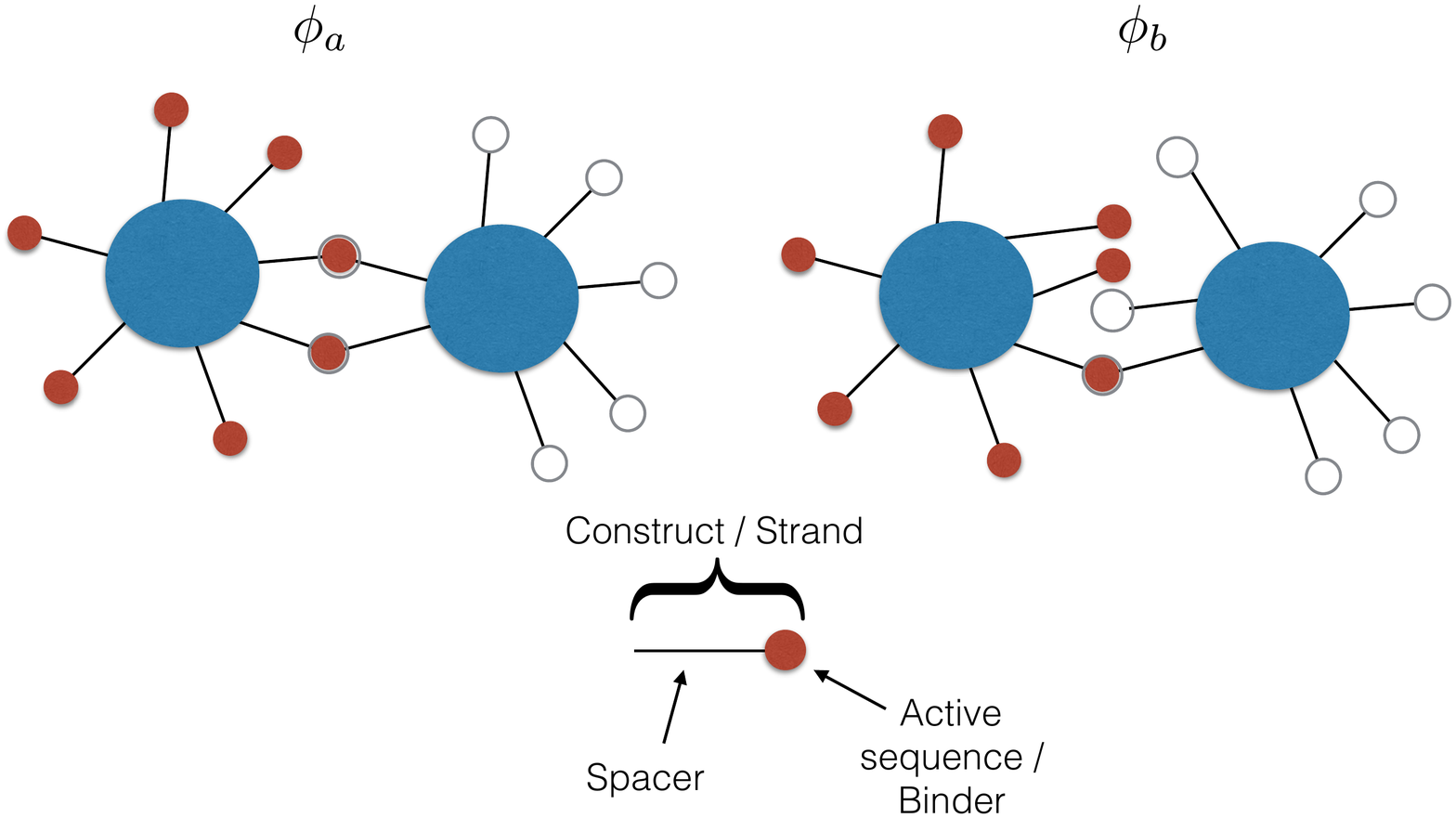}\\
\end{center}
\caption{A schematic representation of a typical DNA-coated colloid, and the working principle behind their induced attraction. In DNACCs, single-stranded DNA, 
here represented as a red or white circle, is grafted onto a colloidal core via a spacer, typically another polymer or double-stranded DNA (dsDNA). Upon recognition of a 
complementary sequence, a single-stranded DNA can hybridise with it to form a joining dsDNA helix, leading to a highly-specific ligand-receptor bond.
The figure presents two possible, mutally exclusive binding configurations, $\phi_a$ and $\phi_b$
Considering all possible 
binding configurations and their energy provides the binding thermodynamics of the system.}\label{fig:system}
\end{figure}

\section{Theory \label{sec:theory}}

\begin{figure}[h!]
\begin{center}
\includegraphics[width=1.0\columnwidth]{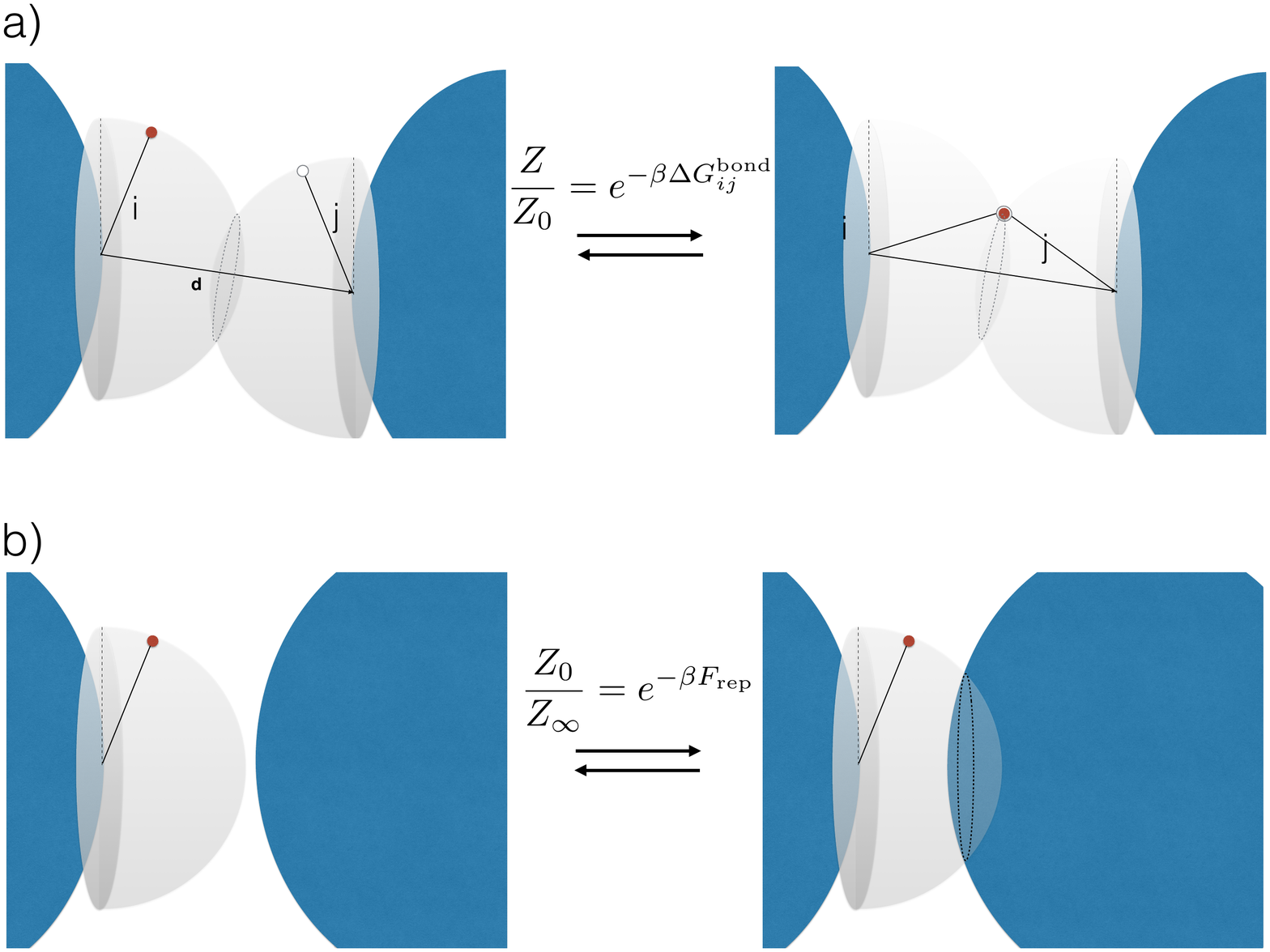}\\
\end{center}
\caption{Schematic representation of the origin of attractive and repulsive forces, here shown for a simple system of strands considered as rigid rods 
and in the case where only a single bond can be formed. a) Upon binding, the system acquires an additional energy $\Delta G^{\rm bond}_{ij}$, which 
depends on both the chemical identity of the active strands (white and red circles) as well as on details of the spacer (black sticks) and position of the 
tethering points (see \eq{single-bond}). In particular, note that whereas in the unbound state (left) the active strands can freely move on a semisphere, 
upon binding (right) they are forced to stay on a circle: This reduction in configurational space is at the origin of $\Delta G^{\rm cnf}$, (see \eq{deltagconfold}). 
b) When all colloids are far apart from each other, the active strands can span a whole semisphere. However, this space is again reduced once colloids 
start to get closer. This reduction is at the origin of the dominant repulsive forces $F_{\rm rep}$ between DNACCs.}
\label{fig:Attractive_and_repulsive}
\end{figure}

The common idea behind basically all theoretical models to describe DNA-mediated interactions in DNACCs \cite{tweezer,schatz,tkachenko-pre,crocker,
melting-theory1,melting-theory2,patrick-jcp,bortolo-pnas,crocker-pnas,miriam,stefano-jcp} is to  account correctly for the interactions induced by the DNA 
strands that coat the colloids (a schematic of the system is depicted in Fig.~\ref{fig:system}). On the one hand, as in general for all colloidal systems stabilised 
by (non-absorbing) grafted polymers, a non-specific repulsion arises from compression of the DNA strands trapped between the colloidal cores. On the other 
hand, a highly specific inter-particle attraction occurs because of the decrease in free-energy due to bond formation \cite{mirkin,alivisatos} (see for reference 
Fig.~\ref{fig:Attractive_and_repulsive}). That bond mediated interactions are the driving force leading to aggregation is also evident from experimental studies 
that show no aggregation between DNACCs grafted with non-complementary strands \cite{mirkin, alivisatos, crocker-pnas}. In fact, other (non-specific) attractive 
interactions, in particular those arising from the van-der-Waals forces between the colloidal cores, are typically much weaker than those arising from bond formation 
on the length-scales at which bond-mediated binding occurs. This is mainly due to the fact that the non-specific steric repulsive forces due to DNA compression prevent 
the colloids to approach at short distances where dispersion forces are strong~\cite{miriam}. For similar reasons, the length-scale for binding, i.e. the typical distance 
between colloidal cores between bound particles, is dictated by the length of the spacer used to graft DNA on the surface of the colloids, which is typically in the range 
of nanometers to tens of nanometers~\cite{tweezer}. 
In various practical realisations of DNACCs, other polymer chains than DNA are grafted to the colloidal surface. 
Using an additional different polymer for steric stabilisation makes it possible to tune repulsion and attraction separately. Both theory and experiments have shown that 
tuning the relative strength of attraction and repulsion can lead to self-assembly of aggregates of different types \cite{stefano-prl,stefano-nature,gang-tuning}.\\
It is crucial to note that the interaction between two grafted colloid is due to both energy and entropy,  
the latter accounting for  the fact that often not just one but several binding configurations are possible (see Fig.\,\ref{fig:system}). Hence, to compute the interaction strength, we need to evaluate a free energy, rather than an energy, as would be the case for atoms in vacuum.
To compute the free energy of interaction of DNACCs, we start from the statistical mechanical relation between the free energy $F$ of a system and its partition function $Z$.
We consider an arbitrary number of colloids, each of which has an arbitrary number of strands grafted on its surface. We choose a reference state where all 
colloids are at infinite distance from each other, and calculate the free-energy with respect to this state:

\begin{align}
\beta F_{\rm interaction} &= - \kbt \log\frac{ Z(\{{\bf R} \})}{ Z^{\infty}} \nonumber \\
						&= - \kbt \log\frac{ Z (\{{\bf R} \})}{Z_0(\{{\bf R} \})} - \kbt \log\frac{Z_0(\{{\bf R} \})}{Z^{\infty}}  \nonumber   \\
						&= F_{\rm att} + F_{\rm rep} 
\label{eq:central}
\end{align}

where $Z=Z_0 + Z_{\rm bound}$ is the partition function of all strands in the system (i.e. a sum over Boltzmann weights), which depends on their 
grafting points $\{{\bf r}\}$, as well as on the positions of all colloids $\{{\bf R} \}$. $Z_0$ is the partial sum over all states where no bond forms, 
$Z_{\rm bound}$ the partial sum over all states with {\it at least} one bond, and finally 
$Z^{\infty}\equiv Z(\{{\bf R = \infty} \})$ is the partition function when all colloids are at infinite separation. With this splitting, we can identify 
two terms, $F_{\rm att}$ and $F_{\rm rep}$, i.e. the attractive and repulsive component of the interaction, respectively. Note that when binding between 
strands on the same particle can occur, this identification cannot be made, although the general approach to calculate $F_{\rm interaction}$ (see Eq.~\ref{eq:central}) still holds.  We stress that no approximation has been made in the derivation of Eq.~\ref{eq:central}. 
Since $Z=Z_0 + Z_{\rm bound}$, $F_{\rm att}$ is always negative (and, hence, attractive). 
Instead $F_{\rm rep}$ is the ratio between two partition functions where bonds are never allowed, although in one case ($Z_0$) strands can feel the 
excluded volume due to all colloids, including the one on which they are grafted, whereas in the other ($Z^{\infty}$) strands can only feel the latter. 
If we assume that only repulsive interactions between the colloidal cores and the strands occur, i.e. we consider non-adsorbing polymers, a good 
approximation for most type of polymers used in DNACCs, then $Z_0 / Z^{\infty} < 1 $, and $F_{\rm rep}$ is always positive (i.e. repulsive).

Use of Eq.\,\ref{eq:central} would allow one to calculate the interaction between DNACCs including all type of many-body interactions arising from bond formations, 
and for arbitrary  position of the colloids. This expression does not assume that DNA-mediated interactions are pairwise additive - and, in fact, often they are not \cite{stefano-prl,
mladek2012quantitative,mladek2}. For this reason, the general expression is needed to compute the interaction free energy of dense phases of DNACCs. \\

Let us next consider how $F_{\rm att}$ and $F_{\rm rep}$ can be calculated.  One route is both obvious and pointless: `brute-force' atomistic simulations. 
Simulations that would describe the colloids, the DNA and the solvent in atomistic detail are simply much too expensive to be used to predict, say, phase diagrams. 
Fully atomistic simulations can certainly be used to study certain aspects of DNACCs, but not to study a suspension containing hundreds or thousands of such 
particles. In what follows, some degree of coarse-graining is therefore essential. Typically, this means that we will use an implicit solvent model and that we will 
assume that the binding free energy between complementary DNA strands is known. In other words: models to describe systems consisting  of many DNACCs 
are always coarse-grained. 

To arrive at an analytically tractable coarse-grained model, the following assumptions are commonly made: 1. we usually ignore the spatial extent of the nucleotide sequences 
that can hybridise to bind to strands together. The hybridising strand is then represented as a point-like attractive site, also called `active site' or `sticky end', tethered 
to the colloidal surface by a long chain -- the spacer or linker \cite{tweezer,tkachenko-pre,crocker,melting-theory1,melting-theory2,patrick-jcp,bortolo-pnas,crocker-pnas,
miriam,stefano-jcp} (see also Fig.\,\ref{fig:system}). Representing the `sticky' binding sequence as a point is a reasonable approximation as long as the spacer is much 
longer than the sticky end, a situation typically encountered in experimental systems \cite{mirkin,alivisatos,crystal-gang,tweezer,melting-theory1}. 
Considering extended sticky ends  results in slightly different binding affinities between constructs \cite{parolini2014thermal,patrick-jcp}. Accounting for the finite spatial extent for sticky 
end does not qualitatively change  the theoretical treatment of DNACCs, and hence we limit our discussion to point-like sticky ends.
2. We ignore steric interactions between different strands and, more generally, any non-selective strand-strand interaction. 
This approximation is very common, but it is likely to break down for dense DNA `brushes'. 

With the above assumptions, $F_{\rm rep}$ can be expressed as:
\eeqq{F_{\rm rep} = -k_{\rm B} T \sum_i  \log \left( \Omega_i / \Omega_i^{\infty}\right),}{eq:frep}
where the index $i$ runs over all strands in the system, and $\Omega_i$ and $\Omega_i^{\infty}$ are, respectively, the partition function of a single linker when the colloids 
are in positions $\{{\bf R} \}$ and the partition function for the case where all colloids are at infinite separation.
For specific geometries and simple polymer models for the linker, such as rigid rods or gaussian chains, there are either exact or very accurate approximate expressions to 
calculate $F_{\rm rep}$ \cite{melting-theory1,miriam,patrick-jcp}. For more complex cases, $F_{\rm rep}$ can be obtained using standard Monte Carlo techniques to 
sample  polymer conformations~\cite{crocker-pnas,bortolo-pnas}. Rogers and Crocker showed that a simple algorithm that assumes that polymers can be described as 
ideal, non-interacting chains that cannot intersect the colloids gave essentially quantitative agreement with the the repulsive interaction determined in experiments~\cite{crocker-pnas}. 
This agreement suggests that in this specific system, the neglect of chain-chain and attractive chain-colloid interactions is justified. In Sec.~\ref{sec:simulations} we discuss 
non-specific interactions between spacers in more detail.

The calculation of $F_{\rm att}$ is more challenging,  since here we need to account for all possible binding configurations with an arbitrary number of bonds 
(see Fig.~\ref{fig:system}). In order to do that, it is useful to split the problem of calculating $F_{\rm att}$ into two parts. The first stage involves calculating the bond 
formation energy for all individual  pairs of $(i,j)$ strands, and the second stage accounts for the fact that there may be many ways in which the DNA strands on 
adjacent colloids may be connected (see for reference Fig.\,\ref{fig:system}).
If the spacer does not influence the structure of the sticky end, one can split the bond formation free energy into two 
terms \cite{melting-theory1,patrick-jcp,stefano-prl,miriam}:

\eeqq{
\Delta G^{\rm bond}_{ij} = \Delta G^0_{ij} + \DGcnfrR.
}{eq:single-bond}

The first term on the r.h.s. of \eq{single-bond} is the  free-energy of binding for untethered, complementary DNA strands $i$ and $j$ in solution:
$\exp\left[-\beta\Delta G^0_{ij}\right] = K^{\rm eq}_{ij} \rho^{\circ}$, where $K^{\rm eq}_{ij}$ denotes the equilibrium binding constant, and $\rho^{\circ}$ 
the standard concentration of 1\,M. The second term represents the configurational entropic cost associated with linking two tethered strands 
$i$ and $j$:  the number of  configurations of a linked $ij$ strand is less than that of the unlinked $i$ and $j$ strands \footnote{A more precise definition
takes into account the Boltzmann-weighted phase space volume of the bound vs unbound states for the linker. As long as one consider non interacting strands,
the two are actually the same}(see Fig.~\ref{fig:Attractive_and_repulsive} {\it a}). 
$\DGcnfrR$ depends on the exact positions of the grafting points, and on the position of all nearby colloids in the system, since these exclude all strand 
configurations that would intersect them. Refs.\,\onlinecite{melting-theory2, patrick-jcp} give an explicit expression that can be used to calculate this 
quantity for arbitrary linkers and colloids positions, which in its most general case reads:

\eeqq{
\DGcnfrR = -k_B T \log \frac{ \Omega_{ij}\left({\bf r}_i,{\bf r}_j,\{{\bf R}_i\}\right)}{\Omega_{i,j}\left({\bf r}_i,{\bf r}_j,\{{\bf R}_i\}\right)\rho^{\circ}},
}{eq:deltagconfold}

where $\Omega_{ij}$ is the partition function for two bound linkers $i$ and $j$, whereas $\Omega_{i,j}$ is the partition function for the unbound case. 
Both terms depend on ${\bf r}_i,{\bf r}_j$ and $\{{\bf R}_i\}$, and can be calculated using coarse-grained models for the linkers \cite{patrick-jcp}.
When strand-strand interactions can be neglected, an equivalent and possibly more illuminating form of \eq{deltagconfold} useful for calculations is:

\eeqq{
\DGcnfrR = -k_B T \log \left(p_{\rm ee} \left( {\bf d}\right) \frac{W_{ij}}{W_i W_j\rho^{\circ}}\right)
}{eq:deltagconf}

where $p_{\rm ee}\left( {\bf d}\right)$ denotes the fraction of all random configurations of the two chains  tethered  at two grafting points at separation 
$\vec{d}$, that have overlapping sticky ends, i.e. it is the probability that two non-interacting theres will form a `bridging' configuration.  The terms 
$W_{ij}$ and $W_{i(j)})$ account for the effect of hard-core overlaps with the colloids: they denote the number of $i$, $j$ or $ij$ configurations that  do not overlap 
with the colloidal cores (see Fig.\ref{fig:Attractive_and_repulsive}). Note that, $W_{i(j)}$ for the unbound linkers is equal to the $\Omega_{i(j)}$s appearing in the 
definition of $F_{\rm rep}$, \eq{frep}.

$\DGcnfrR$ can be either calculated analytically for simple polymer models, or computed via Monte Carlo simulations, as we discuss in Sec.~\ref{sec:simulations}.  
In a small number of (important) cases, $\DGcnf$ can be approximated analytically \cite{melting-theory1,patrick-jcp,tweezer}.
The different theories that have been proposed to estimate $F_{\rm att}$ make different approximations to calculate this quantity. However, all models 
\cite{melting-theory1,melting-theory2,crocker,crocker-pnas} assume that strands are non-interacting (although the interaction can still be included in calculating
the bond energy \cite{patrick-jcp}). This approximation is reasonable when different chains 
are not close to each other.  At higher grafting densities (see e.g.~\cite{mladek2012quantitative,mladek2}), the  simple analytical approaches break down, but the simple theories can still be used to gain insight in the effect of the structure 
of the spacer (e.g. rigid rods or freely jointed chains) on the interactions between DNACCs.
For example, $\DGcnf$ accounts quite well for the experimentally observed  trends  in the melting properties of 
DNACCs~\cite{melting-theory1,melting-theory2}.

Under the assumption that the binding free energy  $\Delta G^{\rm bond}_{ij}$ between $i$ and $j$ does not depend on the presence of other strands (i.e. the non
interaction assumption), knowledge of $\Delta G^{\rm bond}_{ij}$ for all possible pairs of constructs $(i,j)$ is sufficient to write down an exact expression 
for the partition function $Z$:

\eeqq{\frac{Z}{Z_0} = \sum_{\phi} \exp\left[-\sum_{\left(i,j\right) \in \phi} \beta\Delta G^{\rm bond}_{ij}\right]}{eq:fatt}
where  the sum runs over all binding configuration $\phi$ of the system (see Fig.~\ref{fig:system}), 
i.e. $Z/Z_0$ is simply equal to the sum of all Boltzmann weights due to bonds energies over all possible binding configurations. 
Note that $G^{\rm bond}_{ij}$ depends on the position of the strands  via $\Delta G^{\rm cnf}$, see \eq{single-bond}, and hence depends on the spatial configuration of the colloids.

An exact  enumeration of the terms in Eqn.\ref{eq:fatt} becomes rapidly  intractable since the number of 
binding configurations increases very rapidly with the number of strands in the system. Hence Eqn.\ref{eq:fatt}  is typically simplified under 
some assumptions. It is at this level of approximation that differences between the the various analytical theories show up \cite{stefano-jcp,patrick-jcp,crocker-pnas,tweezer,
melting-theory1,tkachenko-pre}. Short of exact enumeration, $Z/Z_0$, and hence $F_{\rm att}$, can be calculated to any given accuracy {\it via} Monte Carlo simulations using thermodynamic integration \cite{frenkel}, providing a route to fully include all possible competitive effects between formation of different binding configurations, as well as about the spatial distribution of the strands. The {\it exact} solution reads\cite{miriam,patrick-jcp,bortolo-pnas}:

\eeqqa{ \beta F_{\rm att}(\beta\Delta G^0) &= \int_{\infty}^{\beta\Delta G^0(T)} \sum_{i<j} p_{ij}(\beta\Delta G^{0'}) d\beta\Delta G^{0'},
}{eq:integration}

where $p_{ij}$ is the probability that a bond between strands $i$ and $j$ is observed, as sampled by MC simulations. 
In Refs.~\onlinecite{stefano-jcp,patrick-jcp} we have shown that the `exact' Monte Carlo results are approximated extremely well by a very simple expression:

\begin{align}
\centering
&F_{\rm att}  = \sum_{i} \log p_i + \frac{1}{2}(1 - p_i),
\label{eq:magic}
\end{align}

and

\begin{align}
\centering
&p_i + \sum_{j} p_i p_j \exp(-\beta \Delta G^{\rm bond}_{ij}) = 1,\qquad \forall\, i,j
\label{eq:magictwo}
\end{align}

where the index $i$ runs over all possible strands in the system. In \eq{magic}, $p_i$ is the probability that strand $i$ is not bound, found by solving the set of
coupled self-consistent equations represented by \eq{magictwo}. Eqns.~\ref{eq:magic} and \ref{eq:magictwo} can be derived using  a recursion formula to count 
all possible binding configurations (see Ref.\,\onlinecite{stefano-jcp}).

In our derivation of \eqtwo{magic}{magictwo} we used an approximate yet very accurate formula for counting the degeneracy of each binding configuration, accounting for the fact that no two linkers can bind to the same target~\cite{stefano-jcp}. From a combinatorial point of view, it turns out that our procedure can be mapped onto Wertheim's theory 
for the equilibrium properties of fluids with highly directional inter-molecular forces \cite{wertheim1,wertheim2,wertheim3}. 
Wertheim's solution is a basic ingredient of the widely used SAFT theory \cite{chapman}. 
In practice, each reactive DNA strand (or, more generally, each binder) must be considered in the Wertheim picture as a {\it spatially  and orientationally fixed} particle with a {\it single binding site}, but each having a {\it unique bond energy}, which in our case must also be defined as in \eq{single-bond}.
In our system, we also have the additional benefit that some of the approximations Wertheim used are always exactly satisfied conditions, in particular the fact that if two strands are bound a third one cannot bind at the same time. Once the problem is cast 
in these terms, it becomes mathematically equivalent to that of Wertheim.

The approximations underlying  \eqtwo{magic}{magictwo} become exact in the case of mobile constructs, see Ref.\,\onlinecite{stefano-prl} and Sec.~\ref{sec:simulations} for more details.
Finally, we stress that although these equations provide the $F_{\rm att}$ induced by the 
interaction between grafted strands. However, it is easy to show that they remain the same when binding between these strands is not direct, but is mediated  by (or compete with),
free complementary strands in solution, another widely used bonding scheme in DNACCs. In practice, \eqtwo{magic}{magictwo} remain valid, as well as the 
configurational part of the bond free-energy $\Delta G^{\rm cnf}$, whereas the solution hybridisation free-energy $\Delta G^0_{ij}$ between strands must be shifted with a density dependent factor.\\
Since \eqtwo{magic}{magictwo} agree extremely well with the corresponding MC simulations, they also properly reproduce all the (important) correlations introduced 
by the competition between strands for the same binding partner. Moreover, they are completely general in that they can be used to treat an arbitrary number of different 
binding pairs, with {\it any} spatial distribution. Hence, they can be used to calculate colloid-colloid interactions also when highly non-homogeneous DNA coatings are present, 
as in the experimental system reported by Pine and coworkers mimicking patchy particles \cite{pine}.\\
Earlier expressions that have been proposed in the literature \cite{tweezer,melting-theory1,melting-theory2,tkachenko-pre} can be derived as approximations of  
\eqtwo{magic}{magictwo} under different assumptions \cite{stefano-jcp}. In particular, the aforementioned models all shared a mean-field approximation for binding, i.e. unlike \eqtwo{magic}{magictwo} they do neglect correlations that are due to the fact that two strands cannot bind simultaneously to the same binding partner  (the  `valence constraint' \cite{patrick-jcp}). Furthermore, 
strands were not only treated as if they were completely independent from each other, but also as if they were all statistically equivalent. The first approximation becomes 
progressively worse as the  binding strength of bonds increases (relative to $k_BT$). 
This drawback can be heuristically corrected by a self--consistent mean field treatment~\cite{bortolo-softmatter}, which yields the mean-field version of \eq{magictwo}. The assumption that all bonds are statistically equivalent gets worse  as the grafting 
of chains becomes more heterogeneous, since in this case each strand will experience a different environment.  However, if we assume that both approximations 
are justified, then the expression for $F_{\rm att}$ simplifies~\cite{tweezer,tkachenko-pre,melting-theory1,melting-theory2}:

\eeqq{F_{\rm att} \approx -k_B T <n>}{eq:approxone}

where $<n>$ is the number of bonds formed in the system. 
This expression for $F_{\rm att}$ is called the ``weak binding'' or the ``Poisson'' approximation~\cite{tweezer,tkachenko-pre,melting-theory1,melting-theory2,crocker-pnas}. 
It provides a lower bound for the real interaction energy, as it always overestimates the number of possible bonds: any correlation would reduce the number of  bonds that can form. 
It has been pointed out \cite{crocker-pnas} that models that are based on the Poisson approximation may overestimate the attractive interaction by as much as two orders of 
magnitude, corresponding to approximately $\approx5 kBT$ per DNA bridge. 
The ``weak binding'' approximation can be obtained formally from the expression given by \eq{integration} by a simple assumption. The integrand in \eq{integration} is nothing but 
the expected number of bound pairs in the system. If this number can be {\it approximated} by a Poisson distribution (an approximation that breaks down at higher binding  
strength \cite{bortolo-pnas}), then \eq{approxone} follows from \eq{integration}. It can be also shown that \eq{approxone} corresponds to a first-order approximation of \eqtwo{magic}{magictwo} (the second-order correction is positive), and its validity is limited to a region where each strand has very few binding partners or small binding energy (hence the 
``weak binding regime'') \cite{stefano-jcp}.
Another possible drawback of \eq{approxone} is that it cannot be generalised easily  to treat an arbitrary number of binding partners with different binding energies, 
since \eq{approxone} was derived considering not just independent, but also equivalent strands. This is a serious limitation, as the phase diagram  of DNACCs can 
be `designed' by making colloids with more than one type of strand (see e.g. Ref.\ \onlinecite{stefano-nature}).

A possible way to estimate $F_{\rm att}$ for an arbitrary number of types of strands was proposed by Rogers and Crocker~\cite{crocker-pnas}. 
Heuristically keeping \eq{approxone} as the expression for the attractive contribution, they proposed to use mass balance equations valid for chemical reactions in solution to describe 
binding between grafted strands, an approach dubbed ``Local Chemical Equilibrium'' (LCE).
In Ref.~\onlinecite{crocker-pnas} it is assumed that the binding  for grafted strands can be treated as an equilibrium binding reaction 
in solution with a non-homogeneous distribution of active sites. This distribution is obtained by Monte Carlo sampling the position of the end-point of polymer chains. 
$F_{\rm att}$ is then approximated as:\\ 

\eeqqa{ C_{\alpha\beta}({\bf r}) &= \frac{C_\alpha({\bf r}) C_\beta ({\bf r})}{\rho^{\circ}} \exp\left(-\beta\Delta G^0_{\alpha\beta}\right) \\
C_\alpha({\bf r}) &\approx C_{\alpha}^0({\bf r}) - \sum_\beta C_{\alpha\beta}({\bf r})\qquad \forall \alpha \\
\beta F_{\rm att} &= \sum_{\alpha<\beta} \int_V d{\bf r}\,\, C_{\alpha\beta}({\bf r } ),
}{eq:n-LCE}

where $C_{\alpha(\beta)}({\bf r}) $ are the equilibrium local concentrations, of free, unbound active sites of type $\alpha(\beta)$ (note that in this formalism all strands 
with the same active site are considered equivalent, whereas in \eqtwo{magic}{magictwo} all strands have different identities) $C_{\alpha(\beta)}^0({\bf r})$, the initial 
concentration of active sites when no binding occurs, and $C_{\alpha\beta}({\bf r}) $ the concentration of bound $\alpha-\beta$ pairs once binding is allowed. 
%%%
 Note that Eq.\,13 represents a very strong constraint: in general, mass conservation makes it valid when concentrations are integrated over the whole 
 volume. Eq.\,13 instead assume that this is true at each point in space, a much stronger statement. In fact, the approximation inEq.\,13  is in general not justified \cite{patrick-jcp}, as we discuss in more detail below.

\subsection{Equilibrium in bulk binary reactions}

In order to derive the (LCE) expression for DNA binding, it is important to realise that tethered DNA strands are distinguishable.
The derivation of the condition for local chemical equilibrium is therefore different from the case where the reactants and products comprise molecules 
that are indistinguishable. As an illustration, we will assume that the tethering of DNA strands results in a confinement of the `reactive' ends, but that the 
free-energy cost of tethering can be ignored (this assumption can subsequently be refined). We consider the hybridisation reaction 
$\alpha+\beta\rightleftharpoons \alpha\beta$. As we can ignore the tethering free energy, we can treat the mixture of $\alpha$, $\beta$ and $\alpha\beta$ 
strands as an ideal gas of {\it distinguishable} particles. 
 We assume that, $\alpha$, $\beta$ and $\alpha\beta$ are confined to a 
volume $V$ that is small, but large enough to contain many molecules.  The partition function of a system with $N_\alpha$ ($N_\beta$) monomers of 
type $\alpha$ ($\beta$) and $N_{\alpha\beta}$ complexes is then:

\begin{align}
Q(N_\alpha,N_\beta,N_{\alpha\beta},V,T) = &{N_\alpha^0!\over N_\alpha! N_{\alpha\beta}!}{N_\beta^0!\over N_\beta! N_{\alpha\beta}!} N_{\alpha\beta}! V^{N^0-N_{\alpha\beta}} \nonumber \\
&\times \prod_{x=\alpha,\beta,\alpha\beta}
\left( \frac{q_{int}^N(x)}{\Lambda_x^{3N}}\right)^{N_x}\;, \label{eq:Qmixture}
\end{align}

where $N^0_\alpha$ ($N^0_\beta$) denotes the original (i.e. pre-reaction) concentration of $\alpha$ ($\beta$) , and $N^0\equiv N^0_\alpha + N^0_\beta$.  
Note that the combinatorial factors in Eqn.~\ref{eq:Qmixture} are not related to indistinguishability but to the number of ways to select $N_{\alpha\beta}$ 
reaction partners from $N_\alpha$ and from $N_\beta$ monomers. The factor $N_{\alpha\beta}!$ accounts for the number of ways in which $N_{\alpha\beta}$ 
distinguishable  particles of type $\alpha$ and the same number of type $\beta$ can be paired. The partition function is a function of $N_{\alpha\beta}$. It is maximised for
\begin{equation}
{N_\alpha N_\beta\over N_{\alpha\beta}V}= {C_\alpha C_\beta\over C_{\alpha\beta}}=\exp\left(-\beta\Delta G^0_{\alpha\beta}\right)/\rho^0
 \;.
\label{eq:equilibrium}
\end{equation}
Hence, for large enough numbers of reactive particles in the volume $V$, the hybridisation of tethered strands obeys the same equilibrium expression as a binary 
gas mixture. However, for small particle numbers this expression fails even qualitatively (see Appendix~\ref{app:A}).  This means that the LCE is reasonable in 
the `mean-field'  limit where any given DNA strand can bind to many others, but it fails badly when this number is $\mathcal{O}$(1) or less. This is typically the case 
when the average distance between grating points is not small compared with the length of the tether. 

It may seem surprising that the approach of Ref.~\onlinecite{crocker-pnas} yields results that were  compatible with the 
experimental data,  given the fact that  Eq.\,12 is  based on an approximation, and that the mean-field assumption is not expected to be particularly good in the 
case studied in Ref.~\onlinecite{crocker-pnas}. It is probably more likely  that the accuracy of the experiments reported in  Ref.~\onlinecite{crocker-pnas} was 
insufficient to discriminate between different models \cite{crocker-answer,patrick-jcp}. 

\begin{figure*}
\vspace{.0cm}
\includegraphics[angle=0,scale=0.55]{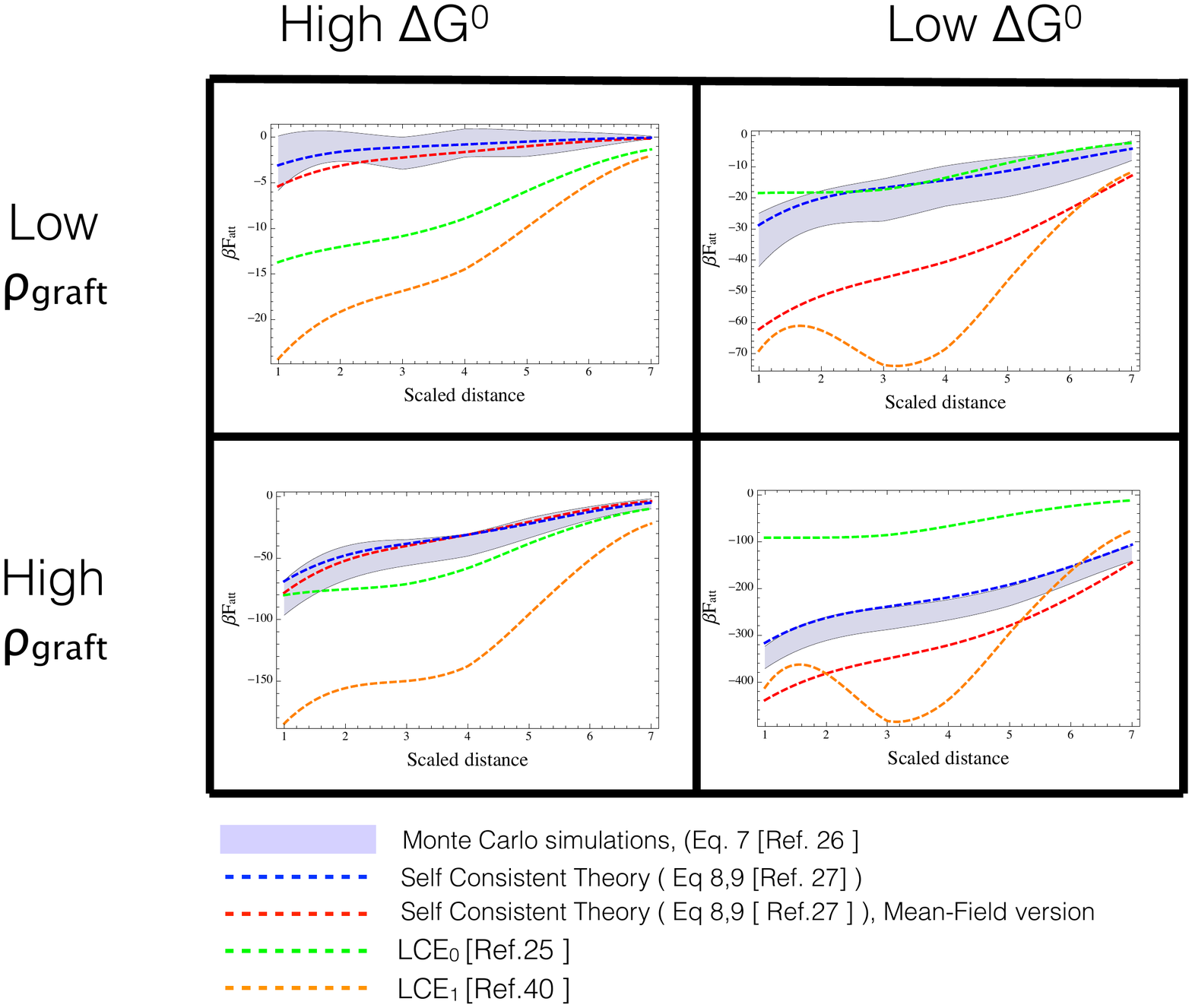}
\vspace{0.cm}
\caption{
$F_{\rm att}$ as a function of distance between DNA-functionalized planes at various temperatures and coating densities, calculated via 
different approximate analytical theories (dashed lines) and Monte Carlo data (shaded region representing the uncertainty determined in the simulations). 
Different panels represent different combinations of $\beta\Delta G^0 = [-10,-15]$ from left to right (roughly proportional to temperature), and strands grafting 
densities $\rho_{\rm graft} \approx [0.006,0.03] \mathrm{strands / nm^2}$ from top to bottom. For reproducibility of these results, we mention that planes have a 
surface area of  $14884\mathrm{nm^2}$ and either 93 or 19 (randomly grafted) strands in the high and low density regime, respectively. 
Color code: Red = Self-consistent theory with added spatial averaging of receptors \cite{patrick-jcp}, Blue = Self-consistent theory, Green = LCE$_0$ 
as in the original version \cite{crocker-pnas}, Orange = LCE$_1$, as modified by Rogers and Manoharan \cite{rogers-manoharan}. Note that only the 
self-consistent theory properly describe the Monte Carlo data in all ranges of densities and $\beta\Delta G^0$. Both forms of LCE provide a decent 
comparison at high $\Delta G^0$ (=high temperatures) and high grafting densities, but fail outside this regime.}
\label{fig:comparison_methods}
\end{figure*}

For this reason, the best way to discriminate between various approximate models is to use computer `experiments' in which `exact' 
Monte Carlo simulations of a specific model are used to compare between different approximations. 
In particular, if a model cannot reproduce the behaviour of a well-characterised model system,   its agreement with experiments must be fortuitous.
In Ref.\,\cite{bortolo-pnas},  we compared  the  different analytical theories with MC calculations  to test the validity of mass balance equations and found 
that LCE can give results that differ substantially from the `exact' simulation results, especially at high binding strengths (see also Fig.\,\ref{fig:comparison_methods}). It should come as no surprise that the agreement worsens at high binding energies as \eq{approxone} is only recovered from our  analytical theory~ \eq{magic} in the weak-binding limit  $p_i\rightarrow 1$~ \cite{stefano-jcp}

A more positive finding is that, in the weak-binding limit \eq{approxone} 
remains still valid for an otherwise arbitrary distribution of binding strengths. No such good news is in store for the original LCE: it fails, even in the weak-binding limit: 
mass-balance equations can only be used to calculate the bond densities when considering the binding between free-strands in solution, not for grafted strands, where the local mass-conservation implicit in Eq.\,12 does not hold.
However, starting from our self-consistent equations, \eqtwo{magic}{magictwo} we can show that a better LCE approximation for the binding densities is given by:

\eeqq{ C_\alpha^0\left({\bf r}\right) = C_\alpha\left({\bf r}\right) / p_\alpha }{eq:correct-LCE}

together with the self-consistent condition:

\eeqq{\int d{\bf r}\,\, C_{\alpha}\left({\bf r}\right) + \sum_\beta C_{\alpha\beta}\left({\bf r}\right) = N_{\alpha},  }{eq:correct2-LCE}

$N_{\alpha}$ being the number of strands of type $\alpha$ in the system.
Once corrected, the LCE treatment gives essentially exact results, at least for weak binding where \eq{approxone} is valid, 
when compared with full Monte Carlo simulations, unless the number of binding partners per linker is very small (see  Appendix A and Fig.~\ref{fig:comparison_methods}). The advantage of \eqtwo{magic}{magictwo} is that these equations agree with the MC simulations over a wide range of binding  regimes, not  just in the weak binding limit (see Fig.\,\ref{fig:comparison_methods}). In this figure, we also show a comparison  with a form of LCE that was recently proposed by Rogers and Manoharan. The latter theory assumes local mass balance but uses an different approximation  for $F_{\rm att}$  (Ref.\,\cite{rogers-manoharan}-SI).\\ 

The theoretical work presented thus far is based on various assumptions, as summarised in the box at the end of this section. 
The validity of the approximations involved in the theory should be tested against numerical simulations for the same model system, whereas the quality of the 
underlying molecular model (flexibility, binding strength, inter-molecular interactions) should be checked against experiments.
 The most stringent experimental test to date is the ability to reproduce the  inter-colloidal  potentials as measured in optical-tweezer experiments. 
Probably even more importantly than reproducing quantitative data, the presented theoretical framework also provides important qualitative insight: it allows one to rationalize various important properties of DNACCs, such as the fact the DNA-mediated binding of DNACCs has a much sharper temperature response than free DNA in solution. As a consequence, DNACCs show higher  selectivity in discriminating between slightly different target DNA strands as used in biosensing applications \cite{taton}. Moreover, the theory explains why DNACCs have
a higher dissociation (`melting') temperature than free DNA in solution \cite{melting-theory1} and it accounts for
the dependence the melting temperature of DNACCs on the type of spacer used for grafting DNA \cite{mirjam-trends,melting-theory1}. 
Within the model, all these effects follow from either multivalent effects due to the large number
of binding configurations, or from the entropic penalty $\Delta G^{\rm cnf}$ arising due to grafting constrains upon bond formation.\\
%%%%%%%%%%%%

\begin{figure}[h!]
%\begin{center}
\includegraphics[width=1.0\columnwidth]{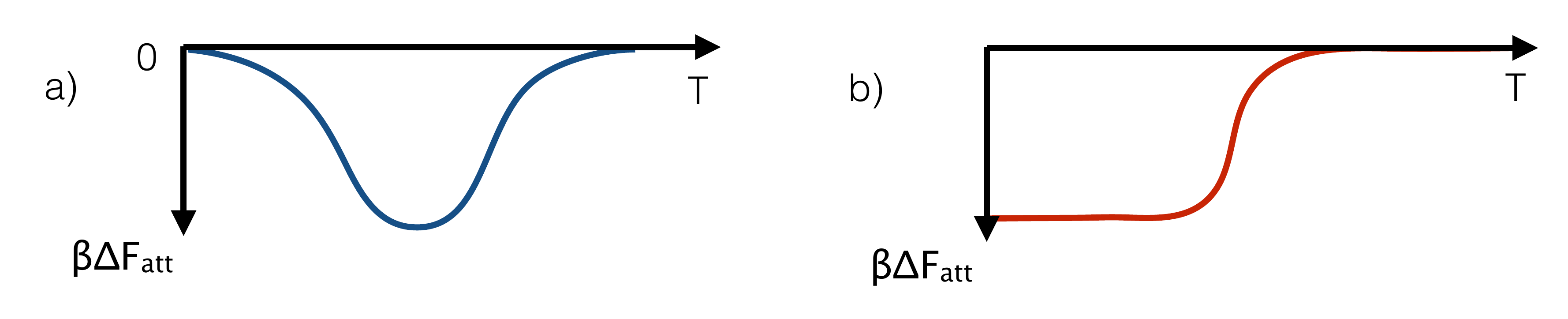}\\
%\end{center}
\caption{ Schematic representation of the pair potential first predicted by theory \cite{stefano-nature} and then supported by experiments and an LCE-based approach \cite{rogers-manoharan}
(see Fig.\, 2,3 in Ref.\ \cite{stefano-nature} and Fig.\,1 in Ref.\,\cite{rogers-manoharan} for the original references.). a) The inter-particle potential is non-monotonic, resulting in an experimental phase diagram with re-entrant melting.
b) The purely entropic origin of the potential at low temperatures provide a temperature-independent interaction (once scaled for $\beta = 1/k_{\rm B} T$) below a certain critical temperature \cite{stefano-nature}. This translates into a temperature-density phase diagram 
with a constant width for the gas-aggregate coexistence region. In Ref.\ \onlinecite{stefano-nature}, the potential were induced via direct hybridisation between grafted strands, whereas in Ref.\ \onlinecite{rogers-manoharan} an indirect hybridisation to free-strands in solution was used.}
\label{fig:compare_mano}
\end{figure}

The availability of an accurate and predictive theory (\eqtwo{magic}{magictwo}) made it possible to explore how the competition between different 
strands for binding would affect the phase behaviour of DNACCs.  Examples are the prediction of the design rules for mixed-strand DNACCs that can undergo re-entrant melting, and the possibility to achieve 
temperature independent interactions due to purely entropy-driven binding~\cite{stefano-nature}. 
Although the subsequent experiments that reported both effects~\cite{rogers-manoharan} used a slightly different system than the example discussed in 
Ref.~\onlinecite{stefano-nature}, , the physical basis of the observed effects was the same.  
%{\color{magenta} {\it remember tu pull off this block if you want to delete linkers}
%(whose discussion can be made using the modified $\Delta G^0_{ij}$, see \eq{free-strands} and Appendix \ref{app:C}), the 
%physical bases of the effects observed were exactly the same}.
%theoretical descriptions of the systems Refs.~\onlinecite{rogers-manoharan,stefano-nature} are essentially the same, even though the details differ.\\
The easiest way to see this is by considering the predicted temperature dependence of the inter-colloidal potential of the systems of 
Refs.~\onlinecite{rogers-manoharan,stefano-nature}. The underlying cause of the reentrant melting is simply that the effective attraction first 
increases with decreasing temperature, and then drops again, whereas the reason for a widened solid-fluid
coexistence stems from purely entropy-driven (and hence temperature independent, considering $F_{\rm att} / k_{{\rm B}T}$) interactions, see Fig.\,\ref{fig:compare_mano}. 
The first behaviour is due to a thermal control of the dominant type of bond in systems featuring competing linkages \cite{stefano-nature,bortolo-softmatter},
while the second arises from purely combinatorial factors between configurations featuring the same number of hybridised 
strands, but a different number of bonds {\it between} colloids (since some of the hybridise strands do not contribute to inter-particle binding) 
\cite{stefano-nature,rogers-manoharan}.

The fact that seemingly different systems can be described by the same theory and exhibit the same behaviour, allows us to make an important general point:
as long as we have quantitative information about  the strength of the ligand-receptor interactions, the nature of the linkers is immaterial. ssDNA is a very convenient molecule to link different colloids together, but the are many other candidates (for example, the cucurbit[8] uril system \cite{cucurbit}, or sialic acid/hemagglutinin complexes when nanocolloids/virus interactions are concerned \cite{haag1,haag2}): the underlying physics  remains exactly the same. We stress this fact because the development of applications non-DNA-based multivalent binding  is expanding rapidly~\cite{multivalency}, and such applications can be rationalized, and designed, using the presented framework.\\
%
%COMMENT ON MIRKIN'S COOPERATIVE THINKING...
%
In the context of DNA-mediated interaction, cooperativity is often mentioned as a crucial aspect. In the theoretical description above, cooperativity plays no role; we assume explicitly that the formation of one bond does not change the binding strength of others,  and yet our theoretical predictions agree very well with experiment. 
To avoid all confusion, we stress that, of course, the fact that one bond between two colloids has formed, makes it easier for subsequent bonds to form, simply because the loss of translational entropy of the binding partner is incurred in the first binding event only. However, that phenomenon has nothing to do with a phenomenon where, at a microscopic level, the strength of subsequent bonds is enhanced by a change in the local environment.
In fact, the latter type of cooperativity is very hard to demonstrate \cite{failure-ercolani,failure-cooperativity2,failure-cooperativity3,failure-cooperativity4}, and it would seem that all experiments where special, cooperative effects were invoked (in the context of an overly simplistic theory)~\cite{failure-ercolani,failure-cooperativity2,failure-cooperativity3,failure-cooperativity4} can, in fact, be explained within the theoretical framework that we discuss here, without cooperativity. This does not mean that there cannot be cooperative effects but simply that the evidence is lacking, if agreement with a simpler theory can be viewed as Occam's razor.

Having said this, there are definitely cases where cooperativity -- but, in this case almost certainly {\it negative} cooperativity -- could play a role: in nano-sized colloids, and possibly for the recently developed micron-size colloids~ \cite{wang2015synthetic},  grafted strands can  be close enough to each other to feel the local change in the environment upon binding,  be it entropic or electrostatic in origin \cite{schatz,delacruz-mirkin,doyen2013dna}. 
Our key message is  that the sharper melting behaviour of DNACCs, the key experimental finding that is often cited as evidence in support of the cooperativity model \cite{schatz}, is perfectly well  explained by our simple model that ignores cooperativity (and, in fact, by even much simpler, non-cooperative models~\cite{geerts-review}). 
On the experimental front, a sharper melting is also clearly observed for systems of micron-sized DNACCs where grafting densities were too low for any cooperativity to occur \cite{crocker,tkachenko-pre}. 
It is appealing that a simple theory that needs not invoke specific, system-dependent cooperative effects can provide a unified  explanation of DNACC melting for 
 a wide range of systems.

%Make a box with all the assumptions
\begin{figure}
\begin{mdframed}
~~~ \\
~~~ \\
{\bf Main approximations implicit in analytical theories}\\
\begin{enumerate}
\item{No strand-strand interaction besides those contained within $\Delta G^{\rm bond}_{ij}$}\\
\item{No interactions besides excluded volume between colloids and strand}\\
\end{enumerate}
\end{mdframed}
\begin{mdframed}
~~~ \\
~~~ \\
{\bf Features included in analytical theories}\\
\begin{enumerate}
\item{Effects induced by changes in binding strength in solution via $\Delta G^0_{ij}$}\\
\item{Effects of spacer configurational entropy, i.e. spacer structure}\\
\item{Presence of multiple bond formation}\\
\item{Competition for bonds between different strands - but only in general treatment of \eqtwo{magic}{magictwo} and equivalent LCE, i.e. effects due to
competing binding configurations}\\
\end{enumerate}
\end{mdframed}
\label{fig:box}
\end{figure}

%\begin{figure}[h!]
%\begin{center}
%\includegraphics[width=0.9\columnwidth]{Approximations.png}\\
%\end{center}
%\end{figure}
%
\section{Simulations \label{sec:simulations}}
Any numerical model of DNACC systems must account for the multi-scale nature of the  problem. 
The reason is that the physical properties of DNACCs depend on both microscopic and mesoscopic features, such as the sticky-end sequences, 
the structure of the spacers, and the grafting density of the DNA units, that enter into the design of the system at different length scales. In fact, the 
differences between various types of DNACCs are substantial. For instance, whilst nano-sized DNACCs may be coated with only a few dozen DNA 
strands, micron-sized DNACCs may be coated with tens of thousands of DNA strands~\cite{wang2015synthetic,wang2015crystallization},  although
typically the coating density in this case is much lower than for nanocolloids. 

Not surprisingly, particles that are that different require different coarse-grained modelling and simulation strategies.
The modelling approaches for micron and nano-sized colloids  will be reviewed in Secs.\ \ref{Micro} and \ref{Nano} respectively. 
In Sec.\ \ref{Mobile} we extend the discussion to micron--sized particles  with mobile rather than grafted strands, e.g. 
suspensions of functionalised lipid--bilayers and emulsions.
Although, in principle, fully atomistic modelling of DNACCs is conceivable, the cost of such an approach would be prohibitive and, 
as in other areas of colloid science, there is no need for such a brute-force approach. However, it should also be clear that microscopic details matter as the interaction between ssDNA strands depends sensitively on their nucleotide sequence. Any model, no matter how coarse-grained, 
%otherwise
 must take this specificity into account.

Roughly speaking, there are three classes of coarse-grained models that account for the DNA specificity in different ways.
The most detailed approaches explicitly account for the pairing of the individually--resolved nucleotides (see part 1 of Fig.\ \ref{Fig_DNA_thermodynamics}) 
\cite{starr2006model,largo2007self,knorowski2012dynamics}.
Recently, more portable but expensive models,  also capturing the double stranded structure of DNA,  
have been introduced \cite{ouldridge2011structural,OXDNAsalt,hinckley2013experimentally}.

%{\color{magenta}  {\it I would delete the following sentence in brackets because the claim would require a longer justification}
%[In this sense, these models can be used to reproduce the shape of the phase diagram, or to predict the stable crystalline structure, but no direct link with the real temperature for phase transitions, nor their kinetics, is present.]
%}

%{\color{magenta} {\it I would delete the following sentence in brackets because it is almost identical to what we say in the nano-section.} [A notable example 
%towards this direction is the work of Lequieu {\it et al.\ } \cite{lequieu2015molecular} and Ding {\it et al.\ } \cite{ding2014insights}, where the 
%SNP3 model was used to study the interparticle potential between nano-colloids functionalised with few DNA strands].}

A more coarse-grained approach uses empirical rules for the binding free-energy of DNA sequences. 
In this description, the binding sequences are represented as structureless points (see part 2 of Fig.\ \ref{Fig_DNA_thermodynamics}),  and the binding 
free-energy is given by \eq{single-bond}, as discussed in  Sec.\,\ref{sec:theory}. 
Finally, a less well-founded, but nevertheless frequently used approach is based on the  `chemical equilibrium'  approximation (see part 3 of Fig.\ \ref{Fig_DNA_thermodynamics}).
The chemical equilibrium theory treats binding reaction as a dimerisation reaction in a confined gas of sticky end groups. The advantage 
of this approach is that it is simple and, when the balance between free and bound ideal constructs is properly 
implemented (see Appendix B in Ref.\ \onlinecite{patrick-jcp}), it becomes equivalent to rigorous treatments. 
However these methods are practical only in the `mean field limit', where the spatial distributions of sticky ends are not strongly dependent on the
exact position of the grafting points.

\subsection{ Micron--sized particles}\label{Micro}
Micron-sized colloids may be coated with several thousands of constructs {\it per} particle 
\cite{crocker-pnas,amorphous1,tweezer,crocker,melting-theory2,melting-theory1,chaikin-subdiffusion,wang2015synthetic,wang2015crystallization}.  
A detailed representation of such a large number of constructs is usually not computationally feasible even when the aim is only to study 
the effective interaction between two colloids. 
The problem is, however, not as serious as it seems, because large particles tend to have a radius of curvature that is much larger 
than the length of the binding strands. Under these conditions, one can use the  Derjaguin approximation (see e.g. \onlinecite{hunter2001foundations})  
to calculate effective pair potentials between colloids \cite{miriam,stefano-nature,walking-colloids} to compute the interaction between two curved 
surfaces from knowledge of the distance-dependence of the interaction between two flat, parallel surfaces. If this approximation is not used, 
a local chemical equilibrium approach (or a ``spatially averaged'' version of the self-consistent equations, see Ref.\,\onlinecite{patrick-jcp,stefano-jcp}, and Fig.\,\ref{fig:comparison_methods}) 
is still tractable, but for all approaches based 
on the explicit calculation of interaction between complementary strands, use of the Derjaguin approximation becomes imperative.
With the Derjaguin approximation, the fact that micron-sized colloids can be coated with very large numbers of DNA strands does not 
create a technical challenge. As discussed Sec.\,\ref{sec:theory}, most theoretical approaches developed to study DNACCs are based on 
the assumption that non-binding strand-strand interactions can be ignored. We stress that this assumption is not essential (and questionable 
for nano-sized colloids: see Ref.~\onlinecite{mladek2,mladek2012quantitative} and Sec.\ \ref{Nano}), but in many cases, the neglect of non-binding strand-strand 
interactions is justified, in which case an accurate estimate of $\Delta G^{\rm cnf}$ is all that is required  to obtain quantitatively accurate 
potentials from computer simulations. 
In practice, we calculate  the  partition function of unbound strands and the configurational part of the binding free energy of two strands, 
$\DG^\mathrm{cnf}$, by simulation. The strand-strand binding free energy $\DG^0$ is not obtained from simulation but from the empirical 
SantaLucia rules~\cite{santalucia,santalucia2004thermodynamics}. Again, more sophisticated rules could be used where necessary.  
Once these two ingredients are known, and if strand-strand interactions can be neglected, the theoretical approaches described in Sec.\,\ref{sec:theory}
can be used to calculate accurate inter-particle potentials.
As was demonstrated in experiments of Bracha for DNA-chips {\it et al.\ }~\cite{bracha2013entropy}
strand conformations are weakly affected by DNA--DNA interaction up to thousand strands {\it per} $\mu$m$^2$. 
Moreover, a rough estimate~\cite{miriam} 
suggests that the contribution of non-bonded spacer interactions to the free energy of the system is of the order of one $k_BT$ unit {\it per} strand, 
which is much smaller in absolute terms than the typical values of the configurational and combinatorial free--energy that are around 
$\approx 10\,$k$_B$T for typical spacers \cite{melting-theory1}.

A notable exception of a micron-size system where non-bonded strand interactions may be significant  is provided by the densely grafted 
micron-sized colloids studied by Pine and co-workers \cite{wang2015synthetic,wang2015crystallization}. The reason why higher grafting densities 
could be reached in Refs. \onlinecite{wang2015synthetic,wang2015crystallization} is that in these experiments, grafting was achieved by using click 
chemistry, rather than via the use of rather bulky streptadivin constructs that  anchor the biotinylated spacers.

%%%%%%%%%%%%
Having sketched the various theoretical approaches to predict the binding strength of DNACCs, we next to consider how to estimate the change in 
the configurational free energy  $\DG^\mathrm{cnf}$  of the grafted strands as the distance between colloids is varied. In general, $\DG^\mathrm{cnf}$  can be computed  using Monte Carlo simulations 
\cite{bortolo-pnas,bortolo-jcp,patrick-jcp}. Practical aspects of these calculations for ssDNA constructs are discussed in Appendix~\ref{app:B}. 
In this modelling, it is important to account for intra-chain Coulomb interactions. For instance, ssDNA constructs have been modelled as 
uniformly charged, freely-jointed chains (FJC) \cite{zhang2001stretching} with screened Debye-H\"{u}ckel interaction (see, e.g.\ Ref.\ \onlinecite{hansen2006theory}). 
Such an approach  has been used to model micron-sized \cite{bortolo-pnas,patrick-jcp,bortolo-jcp} and, as we will discuss later,  nano-sized \cite{mladek2,mladek2012quantitative} DNACCs. When accounting for electrostatic interactions in this way, the electrostatic contribution has to 
be removed from the phenomenological DNA hybridisation free energies, to avoid double counting. 

\begin{figure*}
\vspace{.0cm}
\includegraphics[angle=0,scale=0.55]{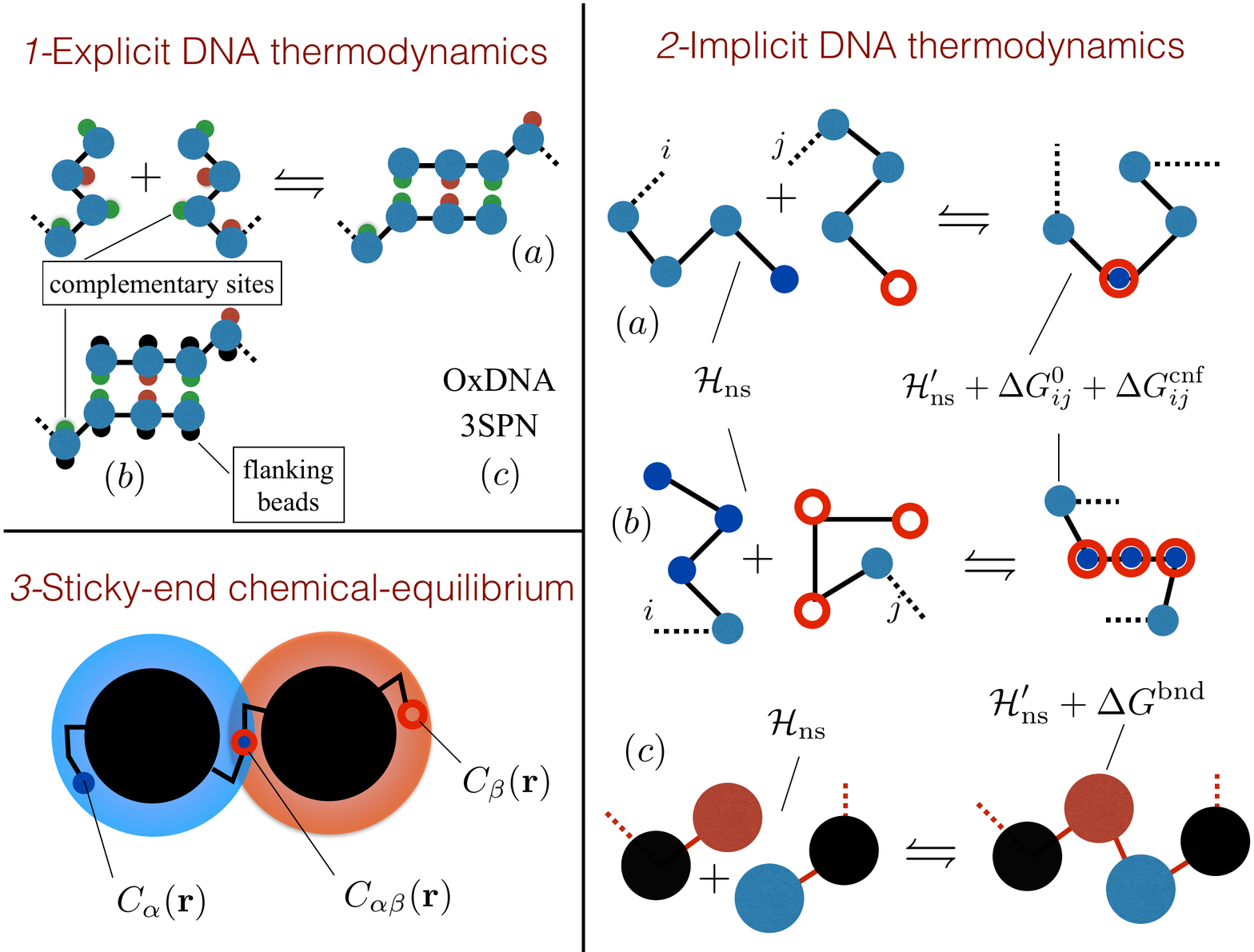} 
\vspace{0.cm}
\caption{ 
Modelling DNA hybridisation in DNACC systems.  1--Models that feature explicit DNA thermodynamics attempt to recover DNA hybridisation 
with an explicit modelling of the dimerisation process. $(a)$ Refs.\ \onlinecite{starr2006model,largo2007self} 
used a bead-and-spring  ssDNA representation decorated by sites that selectively attract. 
$(b)$ This model was ameliorated by Ref.\ \onlinecite{knorowski2012dynamics} that introduced  flanking beads to avoid binding between 
more than two bases. Although the molecular mechanism for binding is retained in these models, none of them tries to match the real thermodynamics
of DNA hybridisation but simply use an effective nucleotide-nucleotide interaction which should be considered as a tunable
model parameter. 
$(c)$ Nucleotide level models, such as SNP3 \cite{hinckley2013experimentally} or OxDNA \cite{ouldridge2011structural,OXDNAsalt}, 
have been parametrised using thermodynamic experiments. 2--Models that feature an implicit DNA hybridisation rely on empirical estimates of the 
hybridisation free energy of the sticky ends ($\Delta G^0_{ij}$) \cite{santalucia,santalucia2004thermodynamics}.
$(a)$ Chains, interacting by non selective Hamiltonians (${\cal H}_\mathrm{ns}$), hybridise by coalescing the two reactive 
end--elements according to  $\Delta G^0_{ij}$ and to a configurational cost ($\Delta G^\mathrm{cnf}_{ij}$), that depends
on the algorithm that generates the hybridised chain \cite{bortolo-pnas,bortolo-jcp,patrick-jcp,mladek2012quantitative}.
Notice that the non selective Hamiltonians of the free and bound states are not the same \cite{mladek2012quantitative,patrick-jcp,bortolo-jcp}.
$(b)$ A similar scheme was previously proposed by Mladek {\it et al. } that also considered the case in which more reactive sites bind resulting 
in a rigid dsDNA segment \cite{mladek2012quantitative,mladek2}. $(c)$ For long $\lambda$--DNA Refs.\ \onlinecite{bozorgui2008liquid,
martinez2010anomalous,francrystals} used a blob representation in which reactive blobs are chained according to an hybridisation free 
energy $\Delta G^\mathrm{bnd}$. 3--For {\it ideal} spacers \cite{patrick-jcp} a local chemical equilibrium balance can be used to calculate the 
fraction of hybridised strands ($\alpha\beta$) \cite{crocker-pnas,patrick-jcp,stefano-prl}. The concentrations of the sticky ends around the 
colloids $C({\bf r})$ are calculated by the end--to--end distribution functions of the tethered spacers \cite{crocker-pnas}, while the equilibrium 
constant between sticky ends is $\Delta G^0_{ij}$.} 
\label{Fig_DNA_thermodynamics}
\end{figure*}
%

%
%\begin{figure}
%\vspace{0.0cm}
%\includegraphics[angle=0,scale=0.3]{Fig_sampling} 
%\vspace{-1.0cm}
%\caption{ 
%\sa{I think we should put this figure in the appendix where we describe the MC calculations, here it's a bit out
%of the blue. If you agree, just move it.}
%Inset bottom: the algorithm used in Refs.\ \onlinecite{bortolo-pnas,patrick-jcp} generates  chains with 
%Rosenbluth weights that are distributed according to a distribution function ($p_0$) which is different from the distribution obtained 
%using equilibrium runs ($p_1$). The average of the Rosenbluth weight using $p_0$ (right $y$--axis), together with the corresponding 
%result for free constructs, is used in Eq.\ \ref{eq:deltagconf} to calculate $\Delta G^\mathrm{cnf}$. 
%Using the overlapping between $p_1$ and $p_0$ and the results of Ref.\ \onlinecite{mooij1994overlapping} we can 
%calculate, in an independent way, $-\log \langle W \rangle_{p_0}$ (left $y$--axis and symbols). The agreement  
%between symbols and the straight line is perfect. The algorithm of Ref.\ \onlinecite{bortolo-jcp} samples between free and 
%bound states in a way that the Rosenbluth weights  are distributed with $p_1$.  In this case $\Delta G^\mathrm{cnf}$ can be 
%measured comparing the time spent by the system in the free and in the bound state \cite{bortolo-jcp}. 
% } \label{Fig_sampling}
%\end{figure}
%
The numerical approach to compute the configurational contribution to the DNACC interaction described in Appendix~\ref{app:B} is efficient
 for short to medium length linkers that can be coarse-grained into at most a few dozen Kuhn segments. However, the same methods would 
 be computationally unfeasible when applied to very long chains. Numerical studies of colloids functionalised by very long DNA strands such as 
$\lambda$-DNA \cite{schmatko2007finite,geerts2008clustering} require a further simplification of the description of the DNA degrees of freedom. 
Bozorgui and Martinez \cite{bozorgui2008liquid,martinez2010anomalous,francrystals} employed a numerical approach based on the `blob' representation 
of long polymer chains\cite{bolhuis2001accurate,pierleoni2007soft} in which  entire DNA constructs are lumped into sites interacting {\it via} gaussian 
potentials \cite{bolhuis2001accurate,pierleoni2007soft,d2012coarse}.  Multi-blob models have been used by Mladek et al.~ \cite{mladek2,mladek2012quantitative}, 
but the approach, whilst fairly accurate, quickly becomes cumbersome. 
The potentials between the blob and the colloids have been parametrised by using explicit `all-monomer' simulations,
while the hybridisation reactions between complementary strands is simulated using a dedicated MC move that attempts to join 
two complementary blobs with a cost equal to $\Delta G^{\rm bond}_{ij}$, where $\Delta G^{\rm cnf}$ is given by the energy of an harmonic
spring connecting the centres of the two blobs of the form $\beta\Delta G^{\rm cnf} = 0.534\left(r/R_{\rm g} - 0.730\right)^2$, 
where $R_{\rm g}$ is the gyration radius of the strand and the numerical values have also been parametrised using all monomer simulations 
(see part 2$c$ of Fig.\ \ref{Fig_DNA_thermodynamics}).

The ultimate aim of coarse-grained modelling of DNACCs is to predict their phase behaviour. Once the tools are in place to compute the 
DNACC interactions, standard free-energy calculations~\cite{frenkel} can be used to compute phase diagrams. Examples of such studies 
can be found in Refs.~\onlinecite{bozorgui2008liquid,martinez2010anomalous,francrystals}.
These simulations showed that DNACCs exhibit very interesting phase behaviour. For example, the simulations showed that the liquid-crystal-vapour 
triple point disappears as the number of grafted chains drops below a threshold of around 12. In that case, the `ground-state' at zero pressure but 
high densities is a liquid, not a crystal~\cite{francrystals}. Recent simulations on `patchy' colloids~\cite{smallenburg2013liquids}, showed 
similar behaviour in such systems. 

%%%%%%%%%%%%%%%
\subsection{Mobile tethers}\label{Mobile}
Recent experiments have studied DNACCs where the DNA strands are mobile, rather than grafted in fixed positions. 
Examples include emulsions \cite{hadorn2012specific, pontani2012biomimetic, feng2013specificity} and lipid bilayers 
\cite{mirjam-mobile,parolini2014thermal,shimobayashi2015direct} functionalised by hydrophobised DNA constructs terminated 
by reactive sticky end sequences. 

It is straightforward to extend the theoretical tools to model DNACCs with grafted DNA strands to the case where these strands are mobile:
due to the translational degrees of freedom of the binders, the hybridisation free--energy needs to include
an extra configurational term ($\Delta G^\mathrm{trn}$) that accounts for the constraint on the relative position of 
two tethering points when bound
 \begin{eqnarray}
 \Delta G^\mathrm{bond}_{ij} = \Delta G^\mathrm{0}_{ij} + \Delta G^\mathrm{cnf}_{ij} + \Delta G^\mathrm{trn}_{ij} \, .
 \label{DGbnd_mobile}
 \end{eqnarray}
 This confining entropic cost $\Delta G^\mathrm{trn}_{ij}$ destabilises DNA  pairing. However, the combinatorial gain 
 is higher in system of mobile tethers because each binder can potentially bind more complementary partners than
in a system with fixed tethering points.  A quantitative analysis of all the entropic and configurational terms of Eq.\ \ref{DGbnd_mobile} 
for a system containing both bridges (i.e. bonds between strands residing on different particles) and loops (i.e. between strands on 
the same particle, when this bears complementary sequences) has been given in Ref.\ \onlinecite{parolini2014thermal}.
 Interestingly, in these systems the self--consistent theory developed in Refs.\ \onlinecite{patrick-jcp, stefano-jcp,bortolo-softmatter} 
(Eqs.\ \ref{eq:magic}, \ref{eq:magictwo}) becomes exact. This was demonstrated in Ref.\ \onlinecite{stefano-prl}, starting from an exact 
 free energy functional $ {\cal F}(\{ n\} , \{ \Delta G^\mathrm{bond} \})$ calculated as a function of the number of bridges between  
 bound colloids ($\{ n \}$) and the matrix of the hybridisation free energies $\Delta G^\mathrm{bond}_{ij}$ (where $i$ and $j$ now represents
 a strand type, not a single specific strand). 
 Taking the large number of binder {\it per } particle  limit, a saddle-point approximation provided the most probable 
 number of bridges between colloids  ($\overline n$)
 \begin{eqnarray}
  {\delta {\cal F}(\{  n\} , \{ \Delta G \}) \over \delta n} \Big{|}_{n=\overline n}  =   0 \, .
 \label{eq:saddle}
 \end{eqnarray}
Remarkably Eq.\ \ref{eq:saddle} is equivalent to Eq.\  \ref{eq:magictwo}. Using $\{ \overline{n} \}$ one can calculate the attractive potential between colloids 
\begin{eqnarray}
F_\mathrm{att} = {\cal F}(\{  \overline{n} \} , \{ \Delta G \})
\label{eq:free}
\end{eqnarray}
and recover exactly the expression given in \ref{eq:magic} (once the number of bridges $\overline n$ are expressed in term 
of probability of making bridges between colloids). It is possible to generalise this discussion to systems featuring different types of bonds 
including not only bridges but also loops \cite{parolini2014thermal}. Slightly different expressions are obtained in presence of infinite 
reservoirs of binders as shown in the study of a vesicle interacting with a large supported lipid bilayer \cite{shimobayashi2015direct}.

Eqs.\ \ref{eq:saddle}, \ref{eq:free} were used to simulate rigid colloids functionalised by mobile strands (a possible experimental 
realisation being that of Ref.\ \onlinecite{mirjam-mobile}) in which colloidal suspensions are sampled using an effective interaction 
free energy $F_\mathrm{att}(\{{\bf R}_i(t)\}) +F_\mathrm{rep}(\{{\bf R}_i(t)\})$ that only depends on the positions of the colloids $\{{\bf R}_i(t)\}$. 

We note that the behaviour of DNACCs with mobile linkers can be qualitatively different from that of DNACCs with grafted linkers, in other 
words: mobile linkers introducs new physics in the system. For example, Ref.\ \onlinecite{stefano-prl} showed that the inter-colloidal potential 
induced by DNA-hybridisation  of mobile linkers is intrinsically a multi--body interaction. This is due to the fact that each construct on a given 
particle can bridge all the neighbouring particles. Hence, different neighbouring colloids may compete for the same linkers. As the number of 
bonds with different neighbours are now correlated,  the free-energy cannot be decomposed into the sum of pair--interactions.
The fact that DNACCs with mobile linkers have intrinsically multi-body interactions can be exploited used to control the `valency' 
(preferred number of complementary neighbours) of a particle \cite{stefano-prl}. This is interesting because conventional approaches to tune the 
valency of colloids are based on the use of different number of interacting `patches'. However, the preparation of well-controlled patchy colloids 
is non-trivial. In contrast, DNACCs with mobile linkers can have a preferred valency without the need of physically creating real patches on the surface
of the colloid~\cite{pine}.

Even more complex effects should be expected for systems of mobile strands where the colloid to which they are attached is deformable, as in the case 
of DNA-functionalised vesicles \cite{parolini2014thermal}, where the flexibility of the lipid membrane plays a non--negligible role \cite{shimobayashi2015direct}
in determining the bond-mediated interaction. For instance, Hu and collaborators \cite{hu2013binding}  considered the interaction between two membranes 
that were coated with complementary proteins (rather than DNA). They found that the equilibrium constant for protein-protein binding 
(i.e. $\Delta G^{\rm bond}$ in our nomenclature) decreases by a factor that is controlled by the roughness of the membranes \cite{hu2013binding}. 
This effect, when translated to DNA-mediated interaction, can be expected to have important consequences for the self-assembly.

% nano 
 \subsection{ Nano--sized particles.} \label{Nano}
 Modelling the interaction between nano-size DNACCs is more demanding than modelling their micron-sized counterparts. The underlying 
 reason is simple: the length-scale separation that justifies the factorisation between the different driving forces and facilitates modelling of micron-sized particles is less pronounced for nano-sized particles.
 
There are other differences as well:  many of the earlier experiments on DNACCs (although not the most recent ones~ \cite{wang2015synthetic,
wang2015crystallization}) used a bulky streptadivin--biotin construct to tether DNA strands to the surface of micron-sized colloids. In contrast, 
DNA anchoring to gold nano-particles is usually achieved by bonding the (much smaller) thiol group to the gold surface.  As a consequence,  
the density of the DNA strands on nano-sized colloids is sufficiently high that the interaction between different DNA--strands cannot be neglected
 for a quantitative modelling.
Constructing a coarse-grained model of nano-sized DNACCs  is therefore much more system-specific than for micron-sized colloids. 
A case in point is the coarse-grained model developed by Mladek {\it et al.\ } in an attempt to reproduce a typical experimental system\cite{crystal-gang}. 
Construction of the model required three coarse-graining steps.
In the first step, simulations were performed on nano-colloids ($12\,$nm diameter) functionalised by 60 homogeneously charged freely-jointed 
chains representing the $\sim$65 base--pair constructs used in experiments \cite{crystal-gang}.  At this level of description, no DNA hybridisation 
was taken into account. Hence, the particles were simply treated as nano-colloids sterically stabilised with densely grafted chains. The simulations 
probed the repulsive interaction between two such colloids. Subsequently, a multi-blob model was constructed to reproduce the computed repulsive 
interaction between the nano-colloids. The ends of the coarse-grained chains were then functionalised with units that could hybridise. However, for 
the nano-sized colloids, the spatial extent of the `sticky' end group could not be ignored, nor the fact that, upon hybridisation, two such groups combine 
to form a rigid double helix that is more rigid than the unbound sticky ends. Hence, although in this model too the hybridisation free energy is calculated 
using the SantaLucia nearest neighbour rules \cite{santalucia,santalucia2004thermodynamics}, the conformational changes upon hybridisation are 
accounted for in the model (see part 2$b$ of Fig.\ \ref{Fig_DNA_thermodynamics}).

Using this fairly complex model, the authors computed the crystallization behaviour of the nano-sized DNACCs and they computed 
the structural properties of the crystalline solid. Encouragingly, the simulations predicted the correct stable crystal structure and gave very 
good estimates for the temperature at which the solid melts. However, the estimated crystal lattice spacing was off by some 10\% 
(presumably because the experimental estimates of the persistence length of ssDNA show a considerable spread). The other important 
finding was that three-body interactions play an important role and must be correctly included in order to predict the structure and melting 
of the crystalline phase. As the above discussion shows, coarse-grained modelling of nano-sized DNACCs is not simple. However, with the 
use of  a coarse-grained polymer model and the empirical SantaLucia nearest neighbour rules, it is still much more efficient than a brute 
force simulation of nano-size DNACCs. Indeed, both the system sizes (hundreds to thousands of colloids) and the timescale for hybridisation are 
such that atomistic simulations of the phase behaviour of DNACCs is out of the question for the foreseeable future, whilst calculation of the 
pair interaction between nano-sized DNACCs is, at lest at present, not an attractive proposition.

There are, however, good reasons to use models for DNA that, whilst not atomistic, due reproduce the structure and dynamics at the nucleotide level. 
In fact, much effort has been devoted to the development of coarse-grained models that are capable of reproducing the hybridisation free energy, 
the binding kinetics  and  the mechanical properties of DNA complexes~\cite{ouldridge2011structural,OXDNAsalt,hinckley2013experimentally}. 
As the interactions that drive DNA hybridisation are primarily the hydrophobic forces between neighbouring bases, such models require a 
nucleotide--level description of the biopolymer.
%HEEERE

% Starr & Sciortino 
Although there have been extensive and very successful simulations of systems with few DNA strands, the models used are still not an 
attractive proposition to simulate the collective behaviour of DNACCs, as such simulations 
would require the study of  systems of hundreds (or thousands) of colloids each coated with dozens of strands.
Moreover, computing the thermodynamic properties from equilibrium simulations requires sampling times that are much longer than the time for a typical 
hybridisation/de-hybridisation events. Hence, at present, this is not yet possible with the models of Refs.~\onlinecite{ouldridge2011structural,OXDNAsalt,hinckley2013experimentally}. 

For this reason, even more coarse-grained -- but still nucleotide-based -- descriptions have been developed starting with the work of 
Starr and Sciortino~\cite{starr2006model,largo2007self}. In the approach of Ref.~\cite{starr2006model,largo2007self}, the nucleotides are 
described as (repulsive) spheres functionalised with a binding site. Neighbouring monomers are bound together by a so-called FENE 
potential \cite{grest1986molecular}.  A bending rigidity term between each three consecutive monomers is also used. 
 A binding site, specifying the type of base, is connected to each monomer {\it via } a FENE potential. These binding sites selectively interact 
 with complementary sites via Lennard Jones potentials. 
This model has been used to study the assembly of dendrimer-like structures functionalised by four DNA arms \cite{starr2006model,largo2007self},  
self--assembly mediated by linkers \cite{hsu2010theoretical} and crystallisation of nanoparticles \cite{dai2010universal}.
% Travesset 
A model that is closely related to that of Starr and Sciortiono \cite{starr2006model} was deployed by Knorowski and collaborators \cite{travesset1}
(see part 1$b$ of Fig.\ \ref{Fig_DNA_thermodynamics}). 
As in Ref.\ \onlinecite{starr2006model,largo2007self}, DNA constructs were modelled by mean of beads carrying selective binding sites.
However, two flanking repulsive sites were added to every bead belonging to the sticky end part of the construct. 
The function of the flanking beads is to  provide directionality to the interaction between complementary bases and to prevent 
(unphysical) multi--base interaction.  
This model has been used to probe the stability of different crystals \cite{travesset1,knorowski2014self} and to investigate the 
dynamics of crystallisation \cite{knorowski2012dynamics}.
% Theodorakis
Recently the model of Starr and Sciortino \cite{starr2006model} has been used by Theodorakis and collaborators to study the pair interactions between 
nanoparticles functionalised by up to eight constructs \cite{theodorakis}. 
This study highlighted the need for a potential further refinement of the model, as the simulations showed the presence of spurious bound 
states in which the ssDNA strands that form a duplex were oriented parallel, rather than anti-parallel. 
Parallel hybridisation is not found in real DNA, as the strongly polarity-dependent Watson-Crick pairing between complementary nucleotides 
uniquely favours anti-parallel hybridised sequences.
 
% de la Cruz 
The model of Knorowski has been further developed  by Li {\it et al.\ } \cite{li2012modeling,li2013thermally}. In particular the size of the beads in the sticky 
part and in the spacer part of the construct have been differentiated to qualitatively account for the different mechanical properties of the ssDNA and of the 
dsDNA spacers. This model is  parametrised in such  a way that each bead corresponds to $\approx\,$2--3 bases.
This relatively detailed model was subsequently used to parametrise an effective pair--potential that is composed of a short range repulsion 
due to the electrostatics of the system, and an attractive interaction provided by the hybridisation of the sticky--ends 
\cite{de-la-cruz2,zwanikken2011local}. Using the effective model the authors rationalised experimental finding on the dynamics of crystallisation \cite{de-la-cruz2,auyeung2014dna}, while the detailed model has been used to probe the stability of the crystals~\cite{li2012modeling,li2013thermally} and to investigate the 
microscopic structure of the facets of DNACCs-made crystal \cite{auyeung2014dna}.
% Schatz
Detailed molecular models have also been used to investigate interaction between DNA constructs and the nanoparticle substrate (e.g.\ Ref.\ \onlinecite{mdDNA,lee2009interaction}), and to study the changes in the structure of dsDNA when used to assemble crystalline structures \cite{ngo2012supercrystals}. However, these studies go beyond the scope of the present review, where we focus on the physical properties of DNACCs that are related to 
the reversible biding of multiple DNA complementary pairs, rather than the formation of a single pair formation.  
\begin{figure*}
\vspace{-2.0cm}
\includegraphics[angle=0,scale=0.6]{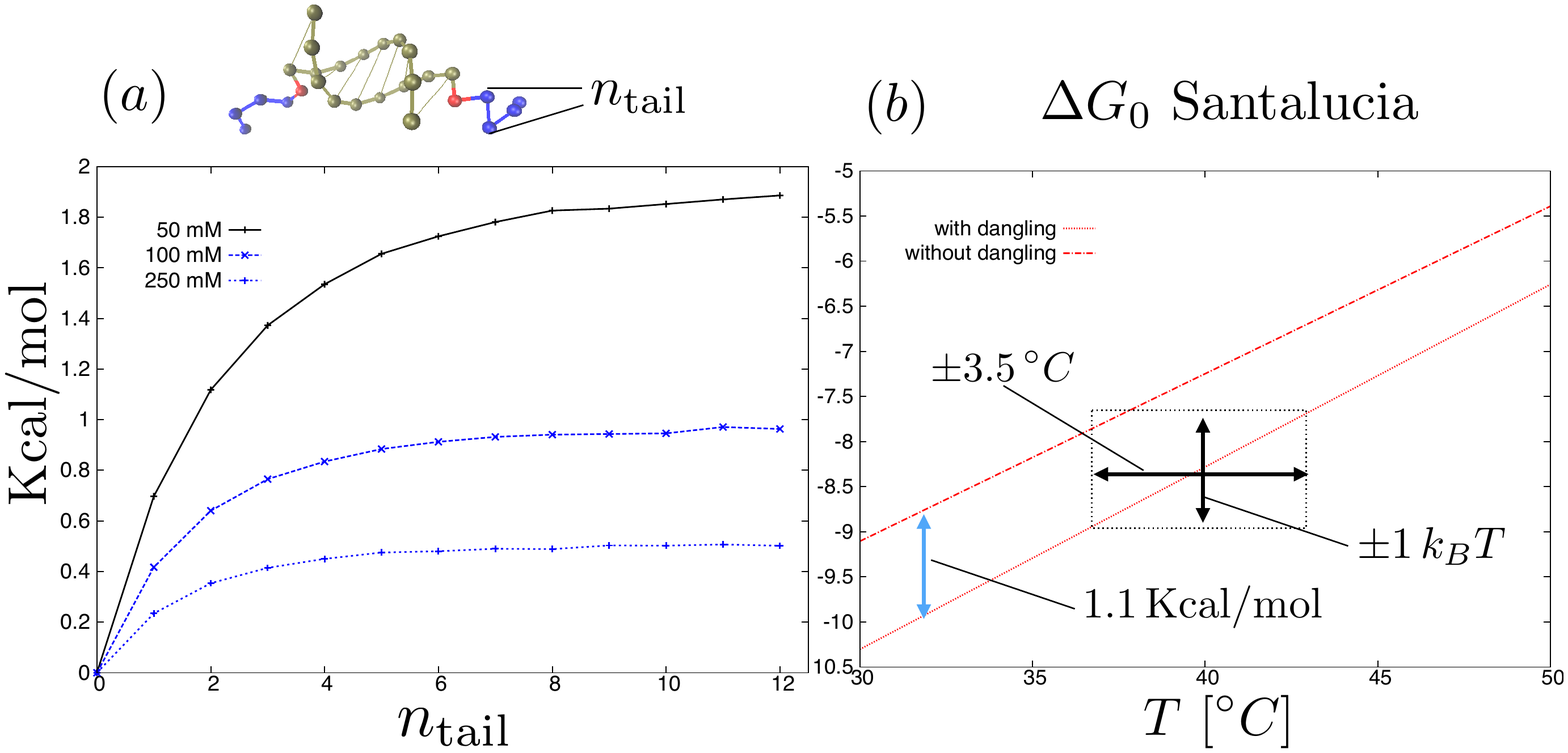} 
\vspace{-2.5cm}
\caption{ $(a)$ Shift in the hybridisation free energy as due to the electrostatic interactions between the inert tails (blue beads in the top 
rendering of the two oligomers) and the paired bases as a function of the number of inert bases.\cite{lorenzo-jacs}  Electrostatic interactions 
are modelled using the linearised Debye--H\"uckel \cite{hansen2006theory} theory at three different salt concentrations. 
Further details of the 
simulation method are reported in Ref.\ \onlinecite{lorenzo-jacs}.
 $(b)$  SantaLucia hybridisation free energy of two  oligomers ($3'$--AACCGACAG--$5'$ and complementary strand) with and without two (G) 
 dangling bases attached to the $5'$ terminals of the reactive sequences for 100$\,$mM NaCl concentration.\cite{santalucia,santalucia2004thermodynamics} 
 The stacking contribution of the Santalucia rules (blue arrow in the figure) is comparable with the tail contribution (see 
 part $a$).  Notice that an error of $\pm 1\, k_B T$ in the nearest-neighbour rules\cite{santalucia,santalucia2004thermodynamics} results in a bigger 
 incertitude of the melting temperature (black arrows). 
 } \label{Fig_tail} 
\end{figure*}

\subsection{ Modelling perspectives } 

There are in principle two ways to construct a proper coarse-grained model for DNACCs. One (the Ôtop-downÕ) approach is to derive a coarse--grained model from a fully atomistic model for DNA\cite{ouldridge2011structural,OXDNAsalt,hinckley2013experimentally,kenward2009brownian}, and use it to simulate DNACCs. Such an approach retains a molecular based description and should be able to describe the hybridisation thermodynamics of DNA. 
The second approach is instead to start with the choice of the ingredients of the coarse-grained model (e.g. ideal chains, rods, monomers, blobs, hybridisation ÔsitesÕ etc) and then to use comparison with the available experimental information to tune its parameters. At present, the Ôtop-downÕ approach is not yet as extensively used as that comparing coarse-grained models directly with experiment.
An encouraging example in this direction is the work of Lequieu {\it et al.\ } \cite{lequieu2015molecular} and Ding {\it et al.\ } \cite{ding2014insights},
who recently reported studies of pair potentials between nanoparticles functionalised by few constructs of reactive DNA, using the 3SPN model 
previously used to study the dimerisation of DNA in solution (Refs.\onlinecite{hinckley2013experimentally,kenward2009brownian} ). 
The OxDNA model of  Ouldridge and coworkers \cite{ouldridge2011structural,OXDNAsalt} is another valuable CG model  
that has been tested in multiple DNA nanotechnology systems. To our knowledge, OxDNA has not yet been used to improve the modelling of DNACCs.

%%%%%%%%

There are other aspects of the modelling that need to be improved. One is related to the naive use of nearest-neighbour rules to compute DNA binding 
free energies in constructs. Recent experimental/theoretical investigation by Di Michele {\it et al.}\ \cite{lorenzo-jacs} showed that the SantaLucia 
nearest-neighbour rules \cite{santalucia,santalucia2004thermodynamics} overestimate the hybridisation free energy $\Delta G^0$ in the case of 
DNA oligomers that are extended to include strings of non-complementary bases~\cite{lorenzo-jacs}.
% more detailed sentence
Considering for instance two complementary sequences of ssDNA decorated by two inert strings at their $5'$ terminals (snapshot of Fig.\ \ref{Fig_tail}$a$), 
Ref.\ \onlinecite{lorenzo-jacs} reported that $\Delta G^0$  increases with the length of the inert tails until reaching a plateau for a number of inert bases 
equal to $\approx 5-7$ (plot in Fig.\ \ref{Fig_tail}$a$). 
In particular the presence of the inert tails destabilises double helix formation and hence decreases the melting temperature of the oligomers. 
 Such an effect is due to the electrostatic interactions between the charged backbones of the DNA that is not included in the nearest neighbour 
 rules \cite{tail_RNA}.
 % RNA
%
Quite surprisingly Ref.\ \onlinecite{lorenzo-jacs} showed that the shift in the hybridisation free energy due to the tails  (at salt concentration that are 
typical of DNA coated colloids experiments \cite{crocker-pnas}) can be as large as the stacking contribution of the nearest-neighbour rules due to the 
first dangling (i.e. non-hybridised) base in the nucleotide sequence, a standard term entering the nearest-neighbour rules (see Fig.\,\ref{Fig_tail}$b$).
The relevance of the inert tails to the calculation of the hybridisation free energy $\Delta G^0$ seems to limit the degree of `responsible' coarse graining 
of DNA mediated interactions  (also in other DNA nanotechnology systems \cite{srinivas2013biophysics}).
In particular if we consider models with explicit hybridisation thermodynamics, the importance of the electrostatic interactions between the inert tails 
and the DNA duplex requires an explicit description of the charged backbones in the tails and in the duplex at the atomistic level.  
Interestingly the latest version of OxDNA introduced Debye-H\"{u}ckel  interactions  \cite{hansen2006theory} between the backbone sites \cite{OXDNAsalt}.

The problem is that the binding free energy computed for a {\it typical} configurations of the dangling segments of the free oligomers\cite{lorenzo-jacs}, 
may not be transferable. This is the case, for instance, if the tails are more stretched when bridging two colloids than when they are free in solution. 
Hence, estimates of $\Delta G^0$ obtained from experiments for free strands might not be reliable once the same strands have been grafted.
These `minor' effects are not so minor when one realises that a bias of one $k_B T$ in $\Delta G^0$ results in a shift in the melting temperatures 
of about $\approx  \pm 3.5\, $K (see Fig.\ \ref{Fig_tail}$b$).  Hence, when analysing accurate experiments~\cite{crocker-pnas}, neglect of the dangling 
tail effects can affect the interpretation of the results, once the experimental resolution gets better than a few Kelvin \cite{tweezer}.

There is also the `no-free-lunch' theorem of coarse graining. As was pointed out in a different context many years ago by Louis, the 
design of an optimal model that carefully reproduces the structural properties of a system {\it may} result in a model that is not representative 
of the thermodynamics of that system (and {\it viceversa}) \cite{louis2002beware}.

Finally we stress something that has been stressed many times: structures may be mechanically stable, but thermodynamically unstable. 
Demonstrating that one structure is more stable than another typically requires the calculation of the free energy (or, more precisely, the 
chemical potential) of both systems. Using MC or Molecular Dynamics to see what kind of structure forms by annealing from a disordered state 
\cite{de-la-cruz2,li2012modeling,li2013thermally,travesset1,knorowski2014self} is interesting, but it does not demonstrate that the structure 
that forms is stable (in fact, Ostwald's rule~\cite{Ostwald1897} suggests the opposite). 
The problem is that the first stages of the self-assembly process are those with a smaller barrier from the starting disordered state. 
These states can simply represent {\it mechanically stable} metastable states, which would transform to the true {\it thermodynamically stable} 
state at much longer times (if at all). These longer times may not be accessible with experiments, and are certainly many orders of magnitude 
larger than the time-scales accessible by computer simulations. We also notice that, in order to correctly reproduce the stable phases observed 
in experiments, studies using this simulated annealing approach typically start with simulations boxes compatible with the shape and number 
of particle experimentally observed, which must thus be known {\it a priori}. Since maximal packing appears to be, at least for conventional 
binary DNACCs system, one of the major driving force behind formation of specific lattices \cite{crystal-mirkin2}, starting with experimentally 
available conditions necessarily biases the simulation toward the expected structure.
In short: constructing a phase diagram requires the calculation of the free energy of each possible 
phase~\cite{frenkel}, unless a direct, hysteresis-free transition between the two phases can be observed. 
One of the main challenges in the simulation of the phase diagram of  DNACCs was precisely the need to carry out free-energy calculations  \cite{martinez2010anomalous,francrystals,mladek2,mladek2012quantitative} that properly account for  the hybridisation between the binders. 
Fortunately, efficient MC algorithms have been developed to calculate the density of states of an aggregate as a function of the number of bridges 
between particles, whose integration can then be used to  calculate the free energy of the crystal~\cite{martinez2010anomalous,francrystals,bortolo-jcp}. 
These algorithms have been successfully combined with various techniques to select among the many possible candidate phases. For example, in view of the rich polymorphism offered by DNACCs, genetic algorithms\cite{gottwald2005predicting,fornleitner2008genetic} have been used to identify possible crystal 
structures, whose relative thermodynamic stability was then probed by free-energy calculations, providing phase diagrams in quantitative agreement 
with experiments without an {\it a priori} knowledge of crystals parameters\cite{mladek2,mladek2012quantitative}.
\section{Conclusions \label{sec:conclusions}}

Theoretical and computational modelling have  proven to be powerful  tools to rationalise the observed behaviour of DNACCs 
\cite{melting-theory1,melting-theory2,tweezer,crocker-pnas,patrick-jcp, de-la-cruz2,li2012modeling,li2013thermally,travesset1,travesset-pnas}, 
and, in some cases, to  predict novel phenomena~\cite{stefano-nature,rogers-manoharan}, with some theoretical predictions for unconventional 
yet experimentally accessible systems still waiting to be tested \cite{francrystals,stefano-prl}.\\
% % 
Understanding DNACCs behaviour means understanding their mutual interactions as a function of the system's parameters. 
In this regard, we have shown how the presently available theoretical framework can be used to design interactions of almost arbitrary 
complexity: How these could lead to interesting effects (see for instance Refs.\ \onlinecite{varrato2012arrested,di2013multistep} for some applications) 
is yet to be systematically investigated. An important factor to highlight in this regard is that the analytical framework we described assumes 
thermodynamic equilibrium for bond formation, i.e. it implicitly considers the bonding network to be equilibrated on the timescale of colloidal diffusion. 
Hence, in order to connect the effective inter-particle potential predicted by theory with experimental observations (that are of finite duration) 
one requires both the bond-formation and bond-breaking rates to be fast compared to the timescale for colloidal diffusion. Until now, this was a 
strong assumption. However, although this was not true for general bonding schemes previously typically employed, Rogers and Manoharan 
\cite{rogers-manoharan} have recently shown a very powerful and general scheme exploiting toe-holding mechanisms that can make such 
assumption essentially correct. Proof of this point is already observable in their experimental results, whose trends could be rationalised using a 
local chemical equilibrium approach that implicitly makes such an assumption. Hence, we would expect such scheme to lead to a more reliable 
comparison between theory and experiments, opening new and truly exciting perspectives for this field.\\ 
% % 
Most of the focus in the DNACCs has probably been on their self-assembly property, and much remains to be done to 
exploit this system to its full potential, in particular in the design of functional materials for applications. In our view, however, other DNACCs based technologies
will benefit from interpreting experiments in view of the presented framework, selective drug delivery and biosensing being two likely candidates.
It should be stressed that, besides describing DNACCs -- systems that have important nanomedicine applications in their own regard 
\cite{taton,mirkin-gene1,mirkin-gene2} -- the theoretical and computational methods outlined here apply more generally to all systems 
whose interactions are dominated by the formation of non-covalent, multiple and reversible ligand-receptor bonds, i.e. supramolecular, 
multivalent based systems.  A  wide variety of such systems are already under intense investigation for nanomedical applications 
\cite{haag1,haag2,kiessling1,kiessling2}, and we envisage important cross-fertilization of ideas between these fields and DNACC research. 
In particular, understanding how to embed binding {\it selectivity} in multivalent-based systems, and not just increase their binding strength, 
seems to be a crucial point where theoretical work \cite{francisco-pnas} might prove useful to rationalise observed effects \cite{kiessling1}, 
and speed up the rational design of applications.\\
% % 
% %
Finally, we  stress that there is much need  for further theoretical and computational developments. For applications that will require 
quantitative predictions, more accurate  experiments are needed to inform the parametrisation of existing, or the development of, novel models. An illustration 
is  the tail effect discussed in Sec.~\ref{sec:simulations}. Understanding the role of non-specific attractive interactions, and also of 
strand-strand interactions, are further challenges that  have still not been properly investigated. The need for such improved modelling has 
become urgent with the arrival of recent experiments of the Pine group, who was able to reach coating densities  that are much higher than 
what was previously achievable~\cite{wang2015synthetic,wang2015crystallization}. Interestingly these experiments reported enhanced 
crystallisation rates; it would be interesting to understand these results using properly extended models.
Finally, other challenges clearly arise when considering systems where the colloid itself can be deformed upon binding (hence effectively 
coupling $\Delta G^{\rm cnf}$ with the shape of the colloidal core), such as the case of functionalised vescicles.\\
% %
After the long gestation period that followed the early work by Mirkin and Alivisatos, the field is moving rapidly. Exciting  times lie ahead of the 
DNACCs field. We hope that this review will have clarified some of the key unifying concepts that need to be understood in the move towards 
the rational design of DNACC-based structures. In this regard, we make freely available a python-based set of routines that we developed to calculate 
ligand-receptor mediated interaction free energies using our self-consistent equations at the following address \url{https://github.com/sangiole/DNACC}.\\

\section*{Additional Material}

Additional data related to this publication is available at the University of Cambridge data repository 
(https://www.repository.cam.ac.uk/handle/1810/xxxxxx).

\section*{Acknowledgements}
This work was supported by ERC AdG grant 227758 (COLSTRUCTION)  and EPSRC Programme Grant EP/I001352/1.
S. A-U. acknowledges financial support from the Beijing Municipal Government Innovation Center for Soft Matter Science and Engineering.
We acknowledge the present and the past members of the group of Daan Frenkel (P.\ Varilly, B.\ Mladeck, 
M.\ E.\ Leunissen, F.\ J.\ Martinez--Veracoechea, J.\ Dobnikar) and of the group of Erika Eiser (L.\ Di Michele, 
T.\ Yanagishima, Z.\ Ruff)  for collaborating with us on the subjects reported by this Perspective.

\appendix

\section{Breakdown of Local Chemical Equilibrium and Mean-Field approximations}\label{app:A}
%\subsection{Equilibrium in bulk binary reactions {\color{magenta} {\it should we use this subsection? there isn't a second}} }
To illustrate the breakdown of LCE and mean--field like approaches in the limit of small numbers of binding strands, first consider the normal dimerisation reaction between two 
``molecules'' $\alpha$ and $\beta$.
\[
\alpha+\beta \rightleftharpoons \alpha\beta \; .
\]
The condition for chemical equilibrium is
\begin{equation}\label{chemeq}
\mu_\alpha+\mu_\beta = \mu_{\alpha\beta}
\end{equation}
We assume that the solutions are ideal. Then the partition
function of species $x$ ($x$=$\alpha$,$\beta$ or $\alpha\beta$) is
\[
Q_x = \frac{V^N}{N!} \frac{q_{int}^N(x)}{\Lambda_x^{3N}}
\]
where $q_{int}(x)$ is the contribution to the partition of $x$
due to all internal degrees of freedom. 
The equilibrium condition \ref{chemeq} can be written as
\begin{equation}\label{partition}
\left(\frac{C_{\alpha\beta}}{C_\alpha\;C_\beta}\right)=
\frac{q_{int}(\alpha\beta)/\Lambda_{\alpha\beta}^3}{q_{int}(\alpha)\;q_{int}(\beta)/(\Lambda_\alpha^3\;\Lambda_\beta^3)}
\end{equation}
where $C_x$ is the number density of species $x$.  To compare
with experiments, it is conventional to express the equilibrium
condition such that the concentrations can be expressed in
Mol/liter.
\[
\frac{[\alpha\beta]}{[\alpha][\beta]}=K
\]
In order to convert to number densities, we note that
$[x]=C_x/\rho^0$. where $\rho^0$ is the ``standard'' number
density (1 molar = 6.022 $\; 10^{26}$ molecules per m$^3$). Then
\[
\left(\frac{C_{\alpha\beta}}{C_\alpha\;C_\beta}\right)=K/\rho^0=\exp\left(-\beta\Delta G^0_{\alpha\beta}\right)/\rho^0
\]
Next consider  the case where we still have an ideal gas mixture, but now with only very few molecules in the `reaction volume' 
(in the case of tethers, this translates into: `a tether of a given type (say $\alpha$) can only bind with a small number of nearby complementary 
tethers.  In that case, LCE breaks down. To see this, consider the extreme (but not unrealistic) case of a volume $V$ containing a single 
molecule of type $\alpha$ and a single copy of $\beta$. As before, $\alpha$ and $\beta$ can react to form $\alpha\beta$. Let us denote 
by $P$ the probability that molecule $\alpha$ (or $\beta$) is part of a dimer, then $P/(1-P)$ is simply given by the ratio of the corresponding 
single-particle partition functions:

\[
\frac{P}{1-P}=\frac{q_{int}(\alpha\beta)/\Lambda_{\alpha\beta}^3}{Vq_{int}(\alpha)\;q_{int}(\beta)/(\Lambda_\alpha^3\;\Lambda_\beta^3)}=K/(V\rho^0)
\]
Now compare this with the LCE expression. The number density of $\alpha$ ($\beta$) equals $P/V$ and the number density of 
$\alpha\beta$ is $(1-P)/V$. Hence, LCE predicts:
\[
\frac{P}{(1-P)^2}=K/(V\rho^0) \;,
\]
which is clearly not correct. Hence, we should not expect LCE and in general Mean Field theories not accounting for the exact position of the tethering points) to work if only 
small numbers of monomers can interact. However, this is precisely 
the situation if the grafting distance between strands is not small compared to the effective length of the strands. 
On the other hand Fig.\ 3 shows that the self--consistent approach performs very well also at low grafting density.

\section{Numerical estimate of configuration contribution to binding free energy}
\label{app:B}
In this Appendix, we describe how the hybridisation free energy between two ssDNA constructs can be been calculated numerically. The method was employed 
in Refs.\ \onlinecite{bortolo-pnas,patrick-jcp}.

\begin{figure}
\vspace{0.0cm}
\includegraphics[angle=0,scale=0.3]{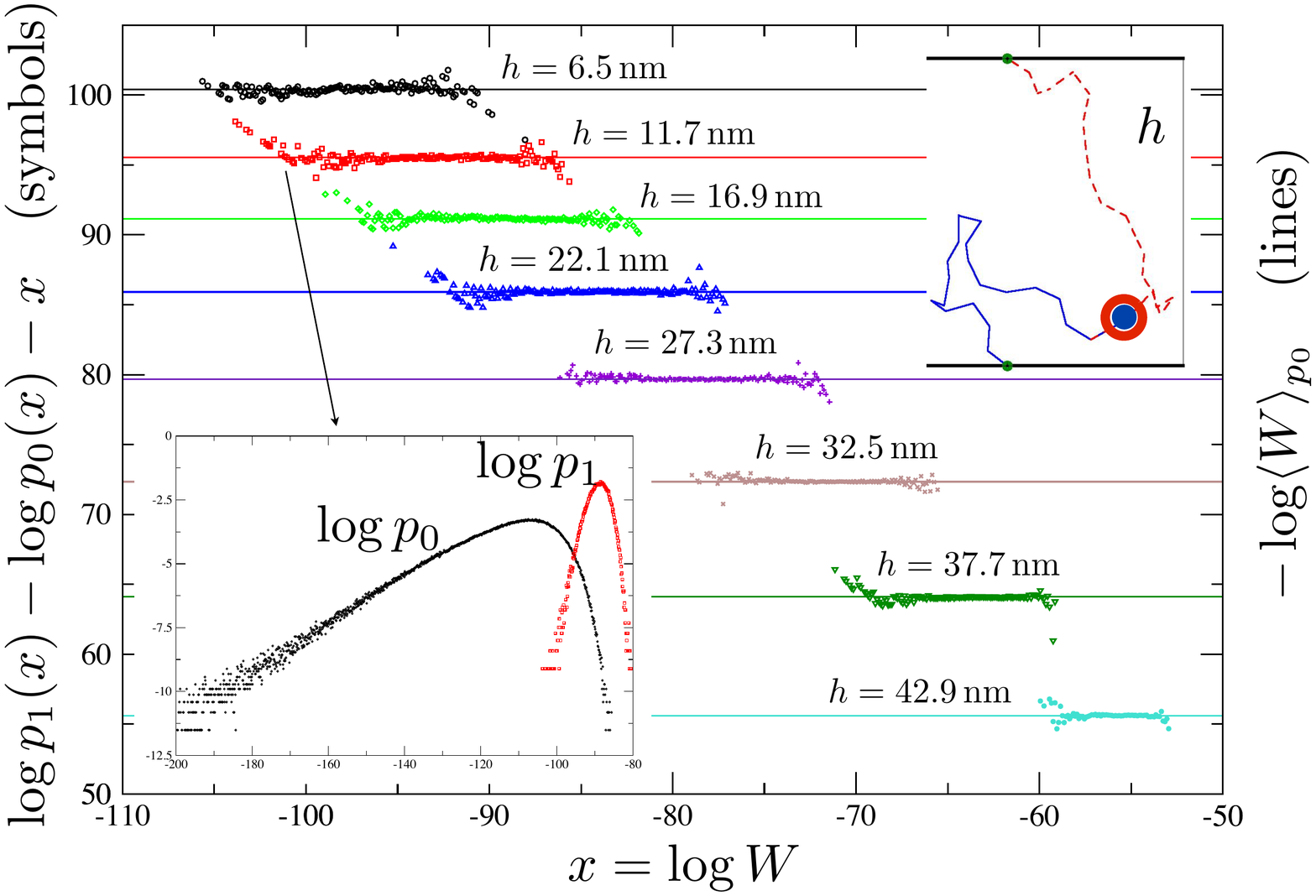} 
\vspace{-1.0cm}
\caption{ Inset bottom: the algorithm used in Refs.\ \onlinecite{bortolo-pnas,patrick-jcp} generates  chains with 
Rosenbluth weights that are distributed according to a distribution function ($p_0$) which is different from the distribution obtained 
using equilibrium runs ($p_1$). The average of the Rosenbluth weight using $p_0$ (right $y$--axis), together with the corresponding 
result for free constructs, is used in Eq.\ \ref{eq:deltagconf} to calculate $\Delta G^\mathrm{cnf}$. 
Using the overlapping between $p_1$ and $p_0$ and the results of Ref.\ \onlinecite{mooij1994overlapping} we can 
calculate, in an independent way, $-\log \langle W \rangle_{p_0}$ (left $y$--axis and symbols). The agreement  
between symbols and the straight line is perfect. The algorithm of Ref.\ \onlinecite{bortolo-jcp} samples between free and 
bound states in a way that the Rosenbluth weights  are distributed with $p_1$.  In this case $\Delta G^\mathrm{cnf}$ can be 
measured comparing the time spent by the system in the free and in the bound state \cite{bortolo-jcp}. 
 } \label{Fig_sampling}
\end{figure}

Two independent Monte Carlo runs are used to generate two interacting  FJCs with free and bound end points, 
respectively, using the Rosenbluth algorithm \cite{frenkel}. 
In the bound case one grows a single longer chain using a bias to constraint the two end--points
to the tethering points \cite{treloar1946statistical,wick2000self}  
 (see the top inset of Fig.\ \ref{Fig_sampling}). 
 In Refs.\ \onlinecite{bortolo-pnas,patrick-jcp}, while growing a hybridised chain, we biased the choice of the new segments using 
 the end-to-end distribution function of an {\it ideal} FJC of length equal to the number of missing segments\cite{treloar1946statistical}
 (notice that this is not the unique choice possible \cite{wick2000self}).
By sampling the corresponding Rosenbluth weights \cite{frenkel} (see the right $y$--axis of Fig.\ \ref{Fig_sampling} for the bound case) 
it is possible to calculate $W_\mathrm{i,j}$  and $W_\mathrm{ij}$ 
\begin{eqnarray}
W_\mathrm{i,j} = \langle W^{( f)} \rangle
& \qquad & 
W_\mathrm{ij} = \langle W^{( b)} \rangle
\end{eqnarray}
(where $W^{( f/ b)}$ are the Rosenbluth weights calculated for free and bound constructs respectively)
that can be used in Eq.\ \ref{eq:deltagconf} to calculate $\Delta G^\mathrm{cnf}$. Notice that in this case 
$p_\mathrm{ee}$ is the end-to-end distribution function of an ideal FJC with the number of segments 
equal to the hybridised construct (see top inset of Fig.\ \ref{Fig_sampling}) and that $W^{( f/ b)}$ also account for 
the non-selective interactions between constructs. 
In the particular case of non--interacting constructs, a further simplification can be used since $W_{i,j}$ factorises into 
the product of two independent Rosenbluth weight ($W_{i,j}=W_i \cdot W_j$), and one can thus use \eq{deltagconf} instead of the 
more general \eq{deltagconfold}.
Notice that for interacting strands ($W_{i,j}\neq W_i \cdot W_j$) a consistent approach would require to include the interactions 
between constructs also when calculating the repulsive part of the potential (see first term of Eq.\ 1).
%{\it Notice we never published pair interactions of interactiong DNA spacers. If you remember at a certain point I considered  such contribution using a virial--like expansion...}
%

%
The accuracy of the aforementioned calculations depends on the quality of the sampling, which can be probed using the overlapping method 
developed in Ref.\ \onlinecite{mooij1994overlapping}, a typical output being shown in Fig.\ \ref{Fig_sampling} (see the caption for further details).
Moreover Ref.\ \onlinecite{bortolo-jcp} developed a dynamic scheme based on the configurational bias method 
\cite{siepmann1992configurational} that allows to sample between hybridised and free chains on the fly, hence without 
the need of precomputing $\DG^\mathrm{cnf}$, for every couple of constructs that could potentially bind.

\bibliographystyle{unsrt}

\begin{thebibliography}{100}

\bibitem{mirkin}
C.~A. Mirkin, R.~C. Letsinger, R.~C. Mucic, and J.~J. Storhoff.
\newblock A {DNA}-based method for rationally assembling nanoparticles into
  macroscopic materials.
\newblock {\em Nature}, 382:607, 1996.

\bibitem{alivisatos}
A.~P. Alivisatos, K.~P. Johnsson, X.~Peng, T.~E. Wilson, C.~J. Loweth, M.~P.
  Bruchez, and P.~G. Schultz.
\newblock Organization of 'nanocrystal molecules' using {DNA}.
\newblock {\em Nature}, 382:609, 1996.

\bibitem{mirkin-sna}
Joshua~I Cutler, Evelyn Auyeung, and Chad~A Mirkin.
\newblock Spherical nucleic acids.
\newblock {\em Journal of the American Chemical Society}, 134(3):1376--1391,
  2012.

\bibitem{crystal-mirkin}
S.~Y. Park, A.~K.~R. Lytton-Jean, B.~Lee, S.~Weigand, G.~C. Schatz, and C.~A.
  Mirkin.
\newblock {DNA}-programmable nanoparticle crystallization.
\newblock {\em Nature Materials}, 451:553--556, 2008.

\bibitem{crystal-gang}
D.~Nykypanchuk, M.~M. Maye, D.~van~der Lelie, and O.~Gang.
\newblock {DNA}-guided crystallization of colloidal nanoparticles.
\newblock {\em Nature Materials}, 451:549--552, 2008.

\bibitem{gang-switch}
Mathew~M Maye, Mudalige~Thilak Kumara, Dmytro Nykypanchuk, William~B Sherman,
  and Oleg Gang.
\newblock Switching binary states of nanoparticle superlattices and dimer
  clusters by dna strands.
\newblock {\em Nature nanotechnology}, 5(2):116--120, 2010.

\bibitem{gang-multiple-particles}
Yugang Zhang, Fang Lu, Kevin~G Yager, Daniel van~der Lelie, and Oleg Gang.
\newblock A general strategy for the dna-mediated self-assembly of functional
  nanoparticles into heterogeneous systems.
\newblock {\em Nature nanotechnology}, 8(11):865--872, 2013.

\bibitem{taton}
T~Andrew Taton, Chad~A Mirkin, and Robert~L Letsinger.
\newblock Scanometric dna array detection with nanoparticle probes.
\newblock {\em Science}, 289(5485):1757--1760, 2000.

\bibitem{mirkin-gene1}
Nathaniel~L Rosi, David~A Giljohann, C~Shad Thaxton, Abigail~KR Lytton-Jean,
  Min~Su Han, and Chad~A Mirkin.
\newblock Oligonucleotide-modified gold nanoparticles for intracellular gene
  regulation.
\newblock {\em Science}, 312(5776):1027--1030, 2006.

\bibitem{mirkin-gene2}
Dan Zheng, David~A Giljohann, David~L Chen, Matthew~D Massich, Xiao-Qi Wang,
  Hristo Iordanov, Chad~A Mirkin, and Amy~S Paller.
\newblock Topical delivery of sirna-based spherical nucleic acid nanoparticle
  conjugates for gene regulation.
\newblock {\em Proceedings of the National Academy of Sciences},
  109(30):11975--11980, 2012.

\bibitem{lorenzo}
Lorenzo Di~Michele and Erika Eiser.
\newblock Developments in understanding and controlling self assembly of
  dna-functionalized colloids.
\newblock {\em Physical Chemistry Chemical Physics}, 15(9):3115--3129, 2013.

\bibitem{geerts-review}
Nienke Geerts and Erika Eiser.
\newblock Dna-functionalized colloids: Physical properties and applications.
\newblock {\em Soft Matter}, 6(19):4647--4660, 2010.

\bibitem{dellago-review}
Panagiotis~E Theodorakis, Nikolaos~G Fytas, Gerhard Kahl, and Christoph
  Dellago.
\newblock Self-assembly of dna-functionalized colloids.
\newblock {\em arXiv preprint arXiv:1503.05384}, 2015.

\bibitem{travesset-review}
C~Knorowski and A~Travesset.
\newblock Materials design by dna programmed self-assembly.
\newblock {\em Current Opinion in Solid State and Materials Science},
  15(6):262--270, 2011.

\bibitem{seeman-mirkin-review}
Matthew~R Jones, Nadrian~C Seeman, and Chad~A Mirkin.
\newblock Programmable materials and the nature of the dna bond.
\newblock {\em Science}, 347(6224):1260901, 2015.

\bibitem{crystal-mirkin2}
R.~Macfarlane, B.~Lee, M.~Jones, N.~Harris, G.~Schatz, and C.~A. Mirkin.
\newblock Nanoparticle superlattice engineering with dna.
\newblock {\em Science}, 334:204--208, 2011.

\bibitem{tweezer}
P.~L. Biancaniello, A.~J. Kim, and J.~C. Crocker.
\newblock Colloidal interactions and self-assembly using {DNA} hybridization.
\newblock {\em Phys. Rev. Lett.}, 94:058302, 2005.

\bibitem{schatz}
Rongchao Jin, Guosheng Wu, Zhi Li, Chad~A Mirkin, and George~C Schatz.
\newblock What controls the melting properties of dna-linked gold nanoparticle
  assemblies?
\newblock {\em Journal of the American Chemical Society}, 125(6):1643--1654,
  2003.

\bibitem{tkachenko-pre}
Nicholas~A Licata and Alexei~V Tkachenko.
\newblock Statistical mechanics of dna-mediated colloidal aggregation.
\newblock {\em Physical Review E}, 74(4):041408, 2006.

\bibitem{crocker}
A.~J. Kim, P.~L. Biancaniello, and J.~C. Crocker.
\newblock Engineering {DNA}-mediated colloidal crystallization.
\newblock {\em Langmuir}, 22(5):1991--2001, 2006.

\bibitem{melting-theory1}
R.~Dreyfus, M.~E. Leunissen, R.~Sha, A.~V. Tkachenko, N.~C. Seeman, D.~J. Pine,
  and P.~M. Chaikin.
\newblock Simple quantitative model for the reversible association of {DNA}
  coated colloids.
\newblock {\em Phys. Rev. Lett.}, 102:048301, 2009.

\bibitem{melting-theory2}
R.~Dreyfus, M.~E. Leunissen, R.~Sha, A.~Tkachenko, N.~C. Seeman, D.~J. Pine,
  and P.~M. Chaikin.
\newblock Aggregation-disaggregation transition of {DNA}-coated colloids:
  Experiments and theory.
\newblock {\em Phys. Rev. E}, 81:041404, 2010.

\bibitem{patrick-jcp}
P.~Varilly, S.~Angioletti-Uberti, B.M. Mognetti, and D.~Frenkel.
\newblock {``A general theory of DNA-mediated and other valence-limited
  colloidal interactions''}.
\newblock {\em The Journal of Chemical Physics}, 137:094108--094122, 2012.

\bibitem{bortolo-pnas}
B.~M. Mognetti, P.~Varilly, { S.} { Angioletti-Uberti}, F.J.
  Martinez-Veracoechea, J.~Dobnikar, M.E. Leunissen, and D.~Frenkel.
\newblock {``Predicting DNA-mediated colloidal pair interactions''}.
\newblock {\em Proceedings of the National Academy of Sciences}, 109, January
  2012.

\bibitem{crocker-pnas}
W.~Benjamin Rogers and John~C. Crocker.
\newblock Direct measurements of dna-mediated colloidal interactions and their
  quantitative modeling.
\newblock {\em Proceedings of the National Academy of Sciences},
  108(38):15687--15692, 2011.

\bibitem{miriam}
M.~E. Leunissen and D.~Frenkel.
\newblock Numerical study of {DNA}-functionalized microparticles and
  nanoparticles: Explicit pair potentials and their implications for phase
  behavior.
\newblock {\em Journal of Chemical Physics}, 134(8):084702, 2011.

\bibitem{stefano-jcp}
{ S.} { Angioletti-Uberti}, P.~Varilly, B.M. Mognetti, A.V. Tkachenko, and
  D.~Frenkel.
\newblock {``Communication: A simple analytical formula for the free energy of
  ligand-receptor-mediated interactions''}.
\newblock {\em The Journal of Chemical Physics}, 138:021102--021106, 2013.

\bibitem{stefano-prl}
{ S.} { Angioletti-Uberti}, P.~Varilly, B.~M. Mognetti, and D.~Frenkel.
\newblock {``Mobile linkers on DNA-coated colloids: valency without patches''}.
\newblock {\em { Physical Review Letters}}, 113:128303--128306, 2014.

\bibitem{stefano-nature}
{ S.} { Angioletti-Uberti}, B.M. Mognetti, and D.~Frenkel.
\newblock { ``Reentrant melting as a design principle for DNA-coated
  colloids''}.
\newblock {\em Nature Materials}, 11:518--522, April 2012.

\bibitem{gang-tuning}
Dmytro Nykypanchuk, Mathew~M. Maye, Daniel van~der Lelie, and Oleg Gang.
\newblock Dna-based approach for interparticle interaction control.
\newblock {\em Langmuir}, 23(11):6305--6314, 2007.

\bibitem{mladek2012quantitative}
Bianca~M Mladek, Julia Fornleitner, Francisco~J Martinez-Veracoechea, Alexandre
  Dawid, and Daan Frenkel.
\newblock Quantitative prediction of the phase diagram of dna-functionalized
  nanosized colloids.
\newblock {\em Physical review letters}, 108(26):268301, 2012.

\bibitem{mladek2}
Bianca~M. Mladek, Julia Fornleitner, Francisco~J. Martinez-Veracoechea,
  Alexandre Dawid, and Daan Frenkel.
\newblock Procedure to construct a multi-scale coarse-grained model of
  dna-coated colloids from experimental data.
\newblock {\em Soft Matter}, 9:7342--7355, 2013.

\bibitem{parolini2014thermal}
Lucia Parolini, Bortolo~M Mognetti, Jurij Kotar, Erika Eiser, Pietro Cicuta,
  and Lorenzo Di~Michele.
\newblock Volume and porosity thermal regulation in lipid mesophases by
  coupling mobile ligands to soft membranes.
\newblock {\em Nature communications}, 6, 2015.

\bibitem{Note1}
A more precise definition takes into account the Boltzmann-weighted phase space
  volume of the bound vs unbound states for the linker. As long as one consider
  non interacting strands, the two are actually the same.

\bibitem{frenkel}
Daan Frenkel and Berend Smit.
\newblock {\em Understanding molecular simulation: from algorithms to
  applications}, volume~1.
\newblock Academic press, 2001.

\bibitem{wertheim1}
M.S. Wertheim.
\newblock Fluids with highly directional attractive forces. i. statistical
  thermodynamics.
\newblock {\em Journal of Statistical Physics}, 35:19--34, 1984.

\bibitem{wertheim2}
M.S. Wertheim.
\newblock Fluids with highly directional attractive forces. ii. thermodynamic
  perturbation theory and integral equations.
\newblock {\em Journal of Statistical Physics}, 35:35--47, 1984.

\bibitem{wertheim3}
M.S. Wertheim.
\newblock Fluids with highly directional attractive forces. iii. multiple
  attraction sites.
\newblock {\em Journal of Statistical Physics}, 42:459--476, 1986.

\bibitem{chapman}
Walter~Gavan Chapman, KE~Gubbins, CG~Joslin, and CG~Gray.
\newblock Theory and simulation of associating liquid mixtures.
\newblock {\em Fluid Phase Equilibria}, 29:337--346, 1986.

\bibitem{pine}
Y.~Wang, Y.~Wang, Dana R., V.~N. Manoharan, L.~Feng, A.~D. Hollingsworth,
  M.~Weck, and D.~J. Pine.
\newblock Colloids with valence and specific directional bonding.
\newblock {\em Nature}, 491:51--55, 2012.

\bibitem{bortolo-softmatter}
B.~Mognetti, M.~E. Leunissen, and D.~Frenkel.
\newblock Controlling the temperature sensitivity of {DNA}-mediated colloidal
  interactions through competing linkages.
\newblock {\em Soft Matter}, 8:2213, 2012.

\bibitem{crocker-answer}
W~Benjamin Rogers and John~C Crocker.
\newblock Reply to mognetti et al.: Dna handshaking interaction data are well
  described by mean-field and molecular models.
\newblock {\em Proceedings of the National Academy of Sciences},
  109(7):E380--E380, 2012.

\bibitem{rogers-manoharan}
W~Benjamin Rogers and Vinothan~N Manoharan.
\newblock Programming colloidal phase transitions with dna strand displacement.
\newblock {\em Science}, 347(6222):639--642, 2015.

\bibitem{mirjam-trends}
M.~E. Leunissen, R.~Dreyfus, R.~Sha, N.~C. Seeman, and P.~M. Chaikin.
\newblock Quantitative study of the association thermodynamics and kinetics of
  {DNA}-coated particles for different functionalization schemes.
\newblock {\em Journal of the American Chemical Society}, 132(6):1903--1913,
  2010.
\newblock PMID: 20095643.

\bibitem{cucurbit}
Urs Rauwald and Oren~A Scherman.
\newblock Supramolecular block copolymers with cucurbit [8] uril in water.
\newblock {\em Angewandte Chemie International Edition}, 47(21):3950--3953,
  2008.

\bibitem{haag1}
Ilona Papp, Christian Sieben, Kai Ludwig, Meike Roskamp, Christoph
  B{\"o}ttcher, Sabine Schlecht, Andreas Herrmann, and Rainer Haag.
\newblock Inhibition of influenza virus infection by multivalent
  sialic-acid-functionalized gold nanoparticles.
\newblock {\em Small}, 6(24):2900--2906, 2010.

\bibitem{haag2}
Jonathan Vonnemann, Christian Sieben, Christopher Wolff, Kai Ludwig, Christoph
  B{\"o}ttcher, Andreas Herrmann, and Rainer Haag.
\newblock Virus inhibition induced by polyvalent nanoparticles of different
  sizes.
\newblock {\em Nanoscale}, 6(4):2353--2360, 2014.

\bibitem{multivalency}
Carlo Fasting, Christoph~A Schalley, Marcus Weber, Oliver Seitz, Stefan Hecht,
  Beate Koksch, Jens Dernedde, Christina Graf, Ernst-Walter Knapp, and Rainer
  Haag.
\newblock Multivalency as a chemical organization and action principle.
\newblock {\em Angewandte Chemie International Edition}, 51(42):10472--10498,
  2012.

\bibitem{failure-ercolani}
Gianfranco Ercolani.
\newblock Assessment of cooperativity in self-assembly.
\newblock {\em Journal of the American Chemical Society}, 125(51):16097--16103,
  2003.

\bibitem{failure-cooperativity2}
Alart Mulder, Jurriaan Huskens, and David~N Reinhoudt.
\newblock Multivalency in supramolecular chemistry and nanofabrication.
\newblock {\em Organic \& biomolecular chemistry}, 2(23):3409--3424, 2004.

\bibitem{failure-cooperativity3}
Jurriaan Huskens, Alart Mulder, Tommaso Auletta, Christian~A Nijhuis, Manon~JW
  Ludden, and David~N Reinhoudt.
\newblock A model for describing the thermodynamics of multivalent host-guest
  interactions at interfaces.
\newblock {\em Journal of the American Chemical Society}, 126(21):6784--6797,
  2004.

\bibitem{failure-cooperativity4}
Francisco~J Martinez-Veracoechea and Mirjam~E Leunissen.
\newblock The entropic impact of tethering, multivalency and dynamic
  recruitment in systems with specific binding groups.
\newblock {\em Soft Matter}, 9(12):3213--3219, 2013.

\bibitem{wang2015synthetic}
Yufeng Wang, Yu~Wang, Xiaolong Zheng, ƒtienne Ducrot, Myung-Goo Lee, Gi-Ra Yi,
  Marcus Weck, and David~J Pine.
\newblock Synthetic strategies toward dna-coated colloids that crystallize.
\newblock {\em Journal of the American Chemical Society}, 137(33):10760--10766,
  2015.

\bibitem{delacruz-mirkin}
Pratik~S Randeria, Matthew~R Jones, Kevin~L Kohlstedt, Resham~J Banga, Monica
  Olvera de~la Cruz, George~C Schatz, and Chad~A Mirkin.
\newblock What controls the hybridization thermodynamics of spherical nucleic
  acids?
\newblock {\em Journal of the American Chemical Society}, 137(10):3486--3489,
  2015.

\bibitem{doyen2013dna}
Matthieu Doyen, Kristin Bartik, and Gilles Bruylants.
\newblock Dna-promoted auto-assembly of gold nanoparticles: Effect of the dna
  sequence on the stability of the assemblies.
\newblock {\em Polymers}, 5(3):1041--1055, 2013.

\bibitem{wang2015crystallization}
Yu~Wang, Yufeng Wang, Xiaolong Zheng, {\'E}tienne Ducrot, Jeremy~S Yodh, Marcus
  Weck, and David~J Pine.
\newblock Crystallization of dna-coated colloids.
\newblock {\em Nature communications}, 6, 2015.

\bibitem{starr2006model}
Francis~W Starr and Francesco Sciortino.
\newblock Model for assembly and gelation of four-armed dna dendrimers.
\newblock {\em Journal of Physics: Condensed Matter}, 18(26):L347, 2006.

\bibitem{largo2007self}
Julio Largo, Francis~W Starr, and Francesco Sciortino.
\newblock Self-assembling dna dendrimers: A numerical study.
\newblock {\em Langmuir}, 23(11):5896--5905, 2007.

\bibitem{knorowski2012dynamics}
Christopher Knorowski and Alex Travesset.
\newblock Dynamics of dna-programmable nanoparticle crystallization: gelation,
  nucleation and topological defects.
\newblock {\em Soft Matter}, 8(48):12053--12059, 2012.

\bibitem{ouldridge2011structural}
Thomas~E Ouldridge, Ard~A Louis, and Jonathan~PK Doye.
\newblock Structural, mechanical, and thermodynamic properties of a
  coarse-grained dna model.
\newblock {\em The Journal of chemical physics}, 134(8):085101, 2011.

\bibitem{OXDNAsalt}
Benedict~EK Snodin, Ferdinando Randisi, Majid Mosayebi, Petr Sulc, John~S
  Schreck, Flavio Romano, Thomas~E Ouldridge, Roman Tsukanov, Eyal Nir, Ard~A
  Louis, et~al.
\newblock Introducing improved structural properties and salt dependence into a
  coarse-grained model of dna.
\newblock {\em arXiv preprint arXiv:1504.00821}, 2015.

\bibitem{hinckley2013experimentally}
Daniel~M Hinckley, Gordon~S Freeman, Jonathan~K Whitmer, and Juan~J de~Pablo.
\newblock An experimentally-informed coarse-grained 3-site-per-nucleotide model
  of dna: Structure, thermodynamics, and dynamics of hybridization.
\newblock {\em The Journal of chemical physics}, 139(14):144903, 2013.

\bibitem{amorphous1}
P.~L. Biancaniello, J.~C. Crocker, D.~A. Hammer, and V.~T. Milam.
\newblock {DNA}-mediated phase behavior of microsphere suspensions.
\newblock {\em Langmuir}, 23(5):2688--2693, 2007.

\bibitem{chaikin-subdiffusion}
Q.~Xu, L.~Feng, R.~Sha, N.~C. Seeman, and P.~M. Chaikin.
\newblock Subdiffusion of a sticky particle on a surface.
\newblock {\em Phys. Rev. Lett.}, 106:228102, Jun 2011.

\bibitem{hunter2001foundations}
Robert~J Hunter.
\newblock Foundations of colloid science.
\newblock 2001.

\bibitem{walking-colloids}
F.J Martinez~Veracoechea, B.M. Mognetti, S.~Angioletti-Uberti, P.~Varilly,
  D.~Frenkel, and J.~Dobnikar.
\newblock {``Designing stimulus-sensitive colloidal walkers''}.
\newblock {\em {Soft Matter}}, 10:3463--3470, 2014.

\bibitem{santalucia}
J.~SantaLucia.
\newblock {A unified view of polymer, dumbbell, and oligonucleotide {DNA}
  nearest-neighbor thermodynamics}.
\newblock {\em Proceedings of the National Academy of Sciences of the United
  States of America}, 95(4):1460--1465, 1998.

\bibitem{santalucia2004thermodynamics}
John SantaLucia~Jr and Donald Hicks.
\newblock The thermodynamics of dna structural motifs.
\newblock {\em Annu. Rev. Biophys. Biomol. Struct.}, 33:415--440, 2004.

\bibitem{bracha2013entropy}
Dan Bracha, Eyal Karzbrun, Gabriel Shemer, Philip~A Pincus, and Roy~H Bar-Ziv.
\newblock Entropy-driven collective interactions in dna brushes on a biochip.
\newblock {\em Proceedings of the National Academy of Sciences},
  110(12):4534--4538, 2013.

\bibitem{bortolo-jcp}
Robin De~Gernier, Tine Curk, Galina~V Dubacheva, Ralf~P Richter, and Bortolo~M
  Mognetti.
\newblock A new configurational bias scheme for sampling supramolecular
  structures.
\newblock {\em The Journal of chemical physics}, 141(24):244909, 2014.

\bibitem{zhang2001stretching}
Yang Zhang, Haijun Zhou, and Zhong-Can Ou-Yang.
\newblock Stretching single-stranded dna: interplay of electrostatic,
  base-pairing, and base-pair stacking interactions.
\newblock {\em Biophysical journal}, 81(2):1133--1143, 2001.

\bibitem{hansen2006theory}
Jean-Pierre Hansen and IR~McDonald.
\newblock {\em Theory of Simple Liquids}.
\newblock Academic Press, 2006.

\bibitem{bozorgui2008liquid}
Behnaz Bozorgui and Daan Frenkel.
\newblock Liquid-vapor transition driven by bond disorder.
\newblock {\em Physical review letters}, 101(4):045701, 2008.

\bibitem{martinez2010anomalous}
Francisco~J Martinez-Veracoechea, Behnaz Bozorgui, and Daan Frenkel.
\newblock Anomalous phase behavior of liquid--vapor phase transition in binary
  mixtures of dna-coated particles.
\newblock {\em Soft Matter}, 6(24):6136--6145, 2010.

\bibitem{francrystals}
F.~J. Martinez-Veracoechea, B.M. Mladek, A.V. Tkachenko, and D.~Frenkel.
\newblock Design rule for colloidal crystals of {DNA}-functionalized particles.
\newblock {\em Phys. Rev. Lett.}, 107:045902, Jul 2011.

\bibitem{schmatko2007finite}
Tatiana Schmatko, Behnaz Bozorgui, Nienke Geerts, Daan Frenkel, Erika Eiser,
  and Wilson~CK Poon.
\newblock A finite-cluster phase in $\lambda$-dna-coated colloids.
\newblock {\em Soft Matter}, 3(6):703--706, 2007.

\bibitem{geerts2008clustering}
Nienke Geerts, Tatiana Schmatko, and Erika Eiser.
\newblock Clustering versus percolation in the assembly of colloids coated with
  long dna.
\newblock {\em Langmuir}, 24(9):5118--5123, 2008.

\bibitem{bolhuis2001accurate}
PG~Bolhuis, AA~Louis, JP~Hansen, and EJ~Meijer.
\newblock Accurate effective pair potentials for polymer solutions.
\newblock {\em The Journal of Chemical Physics}, 114(9):4296--4311, 2001.

\bibitem{pierleoni2007soft}
Carlo Pierleoni, Barbara Capone, and Jean-Pierre Hansen.
\newblock A soft effective segment representation of semidilute polymer
  solutions.
\newblock {\em The Journal of chemical physics}, 127(17):171102, 2007.

\bibitem{d2012coarse}
Giuseppe D'Adamo, Andrea Pelissetto, and Carlo Pierleoni.
\newblock Coarse-graining strategies in polymer solutions.
\newblock {\em Soft Matter}, 8(19):5151--5167, 2012.

\bibitem{smallenburg2013liquids}
Frank Smallenburg and Francesco Sciortino.
\newblock Liquids more stable than crystals in particles with limited valence
  and flexible bonds.
\newblock {\em Nature Physics}, 9(9):554--558, 2013.

\bibitem{hadorn2012specific}
Maik Hadorn, Eva Boenzli, Kristian~T S{\o}rensen, Harold Fellermann,
  Peter~Eggenberger Hotz, and Martin~M Hanczyc.
\newblock Specific and reversible dna-directed self-assembly of oil-in-water
  emulsion droplets.
\newblock {\em Proceedings of the National Academy of Sciences},
  109(50):20320--20325, 2012.

\bibitem{pontani2012biomimetic}
Lea-Laetitia Pontani, Ivane Jorjadze, Virgile Viasnoff, and Jasna Brujic.
\newblock Biomimetic emulsions reveal the effect of mechanical forces on
  cell--cell adhesion.
\newblock {\em Proceedings of the National Academy of Sciences},
  109(25):9839--9844, 2012.

\bibitem{feng2013specificity}
Lang Feng, Lea-Laetitia Pontani, R{\'e}mi Dreyfus, Paul Chaikin, and Jasna
  Brujic.
\newblock Specificity, flexibility and valence of dna bonds guide emulsion
  architecture.
\newblock {\em Soft Matter}, 9(41):9816--9823, 2013.

\bibitem{mirjam-mobile}
Stef A.~J. van~der Meulen and Mirjam~E. Leunissen.
\newblock Solid colloids with surface-mobile dna linkers.
\newblock {\em Journal of the American Chemical Society}, 135(40):15129--15134,
  2013.

\bibitem{shimobayashi2015direct}
SF~Shimobayashi, Bortolo~Matteo Mognetti, Lucia Parolini, Davide Orsi, Pietro
  Cicuta, and Lorenzo Di~Michele.
\newblock Direct measurement of dna-mediated adhesion between lipid bilayers.
\newblock {\em Physical Chemistry Chemical Physics}, 2015.

\bibitem{hu2013binding}
Jinglei Hu, Reinhard Lipowsky, and Thomas~R Weikl.
\newblock Binding constants of membrane-anchored receptors and ligands depend
  strongly on the nanoscale roughness of membranes.
\newblock {\em Proceedings of the National Academy of Sciences},
  110(38):15283--15288, 2013.

\bibitem{grest1986molecular}
Gary~S Grest and Kurt Kremer.
\newblock Molecular dynamics simulation for polymers in the presence of a heat
  bath.
\newblock {\em Physical Review A}, 33(5):3628, 1986.

\bibitem{hsu2010theoretical}
Chia~Wei Hsu, Francesco Sciortino, and Francis~W Starr.
\newblock Theoretical description of a dna-linked nanoparticle self-assembly.
\newblock {\em Physical review letters}, 105(5):055502, 2010.

\bibitem{dai2010universal}
Wei Dai, Sanat~K Kumar, and Francis~W Starr.
\newblock Universal two-step crystallization of dna-functionalized
  nanoparticles.
\newblock {\em Soft Matter}, 6(24):6130--6135, 2010.

\bibitem{travesset1}
C.~Knorowski, S.~Burleigh, and A.~Travesset.
\newblock Dynamics and statics of dna-programmable nanoparticle self-assembly
  and crystallization.
\newblock {\em Phys. Rev. Lett.}, 106:215501, May 2011.

\bibitem{knorowski2014self}
Christopher Knorowski and Alex Travesset.
\newblock Self-assembly and crystallization of hairy (f-star) and dna-grafted
  nanocubes.
\newblock {\em Journal of the American Chemical Society}, 136(2):653--659,
  2014.

\bibitem{theodorakis}
Panagiotis~E. Theodorakis, Christoph Dellago, and Gerhard Kahl.
\newblock A coarse-grained model for dna-functionalized spherical colloids,
  revisited: Effective pair potential from parallel replica simulations.
\newblock {\em The Journal of Chemical Physics}, 138(2):--, 2013.

\bibitem{li2012modeling}
Ting~ING Li, Rastko Sknepnek, Robert~J Macfarlane, Chad~A Mirkin, and Monica
  Olvera de~la Cruz.
\newblock Modeling the crystallization of spherical nucleic acid nanoparticle
  conjugates with molecular dynamics simulations.
\newblock {\em Nano letters}, 12(5):2509--2514, 2012.

\bibitem{li2013thermally}
Ting~ING Li, Rastko Sknepnek, and Monica Olvera de~la Cruz.
\newblock Thermally active hybridization drives the crystallization of
  dna-functionalized nanoparticles.
\newblock {\em Journal of the American Chemical Society}, 135(23):8535--8541,
  2013.

\bibitem{de-la-cruz2}
Subas Dhakal, Kevin~L Kohlstedt, George~C Schatz, Chad~A Mirkin, and Monica
  Olvera de~la Cruz.
\newblock Growth dynamics for dna-guided nanoparticle crystallization.
\newblock {\em ACS nano}, 7(12):10948--10959, 2013.

\bibitem{zwanikken2011local}
Jos~W Zwanikken, Peijun Guo, Chad~A Mirkin, and Monica Olvera de~la Cruz.
\newblock Local ionic environment around polyvalent nucleic acid-functionalized
  nanoparticles.
\newblock {\em The Journal of Physical Chemistry C}, 115(33):16368--16373,
  2011.

\bibitem{auyeung2014dna}
Evelyn Auyeung, Ting~ING Li, Andrew~J Senesi, Abrin~L Schmucker, Bridget~C
  Pals, Monica~Olvera de~La~Cruz, and Chad~A Mirkin.
\newblock Dna-mediated nanoparticle crystallization into wulff polyhedra.
\newblock {\em Nature}, 505(7481):73--77, 2014.

\bibitem{mdDNA}
O.~Lee and G.~C. Schatz.
\newblock Molecular dynamics simulation of {DNA}-functionalized gold
  nanoparticles.
\newblock {\em J. Phys. Chem. C}, 113:2316--2321, 2009.

\bibitem{lee2009interaction}
One-Sun Lee and George~C Schatz.
\newblock Interaction between dnas on a gold surface.
\newblock {\em The Journal of Physical Chemistry C}, 113(36):15941--15947,
  2009.

\bibitem{ngo2012supercrystals}
Van~A Ngo, Rajiv~K Kalia, Aiichiro Nakano, and Priya Vashishta.
\newblock Supercrystals of dna-functionalized gold nanoparticles: a
  million-atom molecular dynamics simulation study.
\newblock {\em The Journal of Physical Chemistry C}, 116(36):19579--19585,
  2012.

\bibitem{lorenzo-jacs}
Lorenzo Di~Michele, Bortolo~M. Mognetti, Taiki Yanagishima, Patrick Varilly,
  Zachary Ruff, Daan Frenkel, and Erika Eiser.
\newblock Effect of inert tails on the thermodynamics of dna hybridization.
\newblock {\em Journal of the American Chemical Society}, 136(18):6538--6541,
  2014.

\bibitem{kenward2009brownian}
Martin Kenward and Kevin~D Dorfman.
\newblock Brownian dynamics simulations of single-stranded dna hairpins.
\newblock {\em The Journal of chemical physics}, 130(9):095101, 2009.

\bibitem{lequieu2015molecular}
Joshua~P Lequieu, Daniel~M Hinckley, and Juan~J de~Pablo.
\newblock A molecular view of dna-conjugated nanoparticle association energies.
\newblock {\em Soft Matter}, 2015.

\bibitem{ding2014insights}
Yajun Ding and Jeetain Mittal.
\newblock Insights into dna-mediated interparticle interactions from a
  coarse-grained model.
\newblock {\em The Journal of chemical physics}, 141(18):184901, 2014.

\bibitem{tail_RNA}
It has been reported that dangling sequences can stabilise RNA-RNA duplexes
  \cite{ohmichi2002long,limmer19933}. This result may look puzzling in view of
  the conclusions of Ref.\ \onlinecite{lorenzo-jacs}. It is likely that tails
  stabilise RNA duplexes by selective interactions as suggested by the fact
  that, contrarily to Ref.\ \onlinecite{lorenzo-jacs}, the effect is
  sequence-dependent\cite{ohmichi2002long,limmer19933}.

\bibitem{srinivas2013biophysics}
Niranjan Srinivas, Thomas~E Ouldridge, Petr {\v{S}}ulc, Joseph~M Schaeffer,
  Bernard Yurke, Ard~A Louis, Jonathan~PK Doye, and Erik Winfree.
\newblock On the biophysics and kinetics of toehold-mediated dna strand
  displacement.
\newblock {\em Nucleic acids research}, 41(22):10641--10658, 2013.

\bibitem{louis2002beware}
AA~Louis.
\newblock Beware of density dependent pair potentials.
\newblock {\em Journal of Physics: Condensed Matter}, 14(40):9187, 2002.

\bibitem{Ostwald1897}
Wilhelm Ostwald.
\newblock Studien \"{u}ber die bildung und umwandlung fester k\"{o}rper.
\newblock {\em Z. Phys. Chem.}, 22:289--220, 1897.

\bibitem{gottwald2005predicting}
Dieter Gottwald, Gerhard Kahl, and Christos~N Likos.
\newblock Predicting equilibrium structures in freezing processes.
\newblock {\em The Journal of chemical physics}, 122(20):204503, 2005.

\bibitem{fornleitner2008genetic}
Julia Fornleitner, Federica~Lo Verso, Gerhard Kahl, and Christos~N Likos.
\newblock Genetic algorithms predict formation of exotic ordered configurations
  for two-component dipolar monolayers.
\newblock {\em Soft Matter}, 4(3):480--484, 2008.

\bibitem{travesset-pnas}
Alex Travesset.
\newblock Binary nanoparticle superlattices of soft-particle systems.
\newblock {\em Proceedings of the National Academy of Sciences},
  112(31):9563--9567, 2015.

\bibitem{varrato2012arrested}
Francesco Varrato, Lorenzo Di~Michele, Maxim Belushkin, Nicolas Dorsaz, Simon~H
  Nathan, Erika Eiser, and Giuseppe Foffi.
\newblock Arrested demixing opens route to bigels.
\newblock {\em Proceedings of the National Academy of Sciences},
  109(47):19155--19160, 2012.

\bibitem{di2013multistep}
Lorenzo Di~Michele, Francesco Varrato, Jurij Kotar, Simon~H Nathan, Giuseppe
  Foffi, and Erika Eiser.
\newblock Multistep kinetic self-assembly of dna-coated colloids.
\newblock {\em Nature communications}, 4, 2013.

\bibitem{kiessling1}
Coby~B Carlson, Patricia Mowery, Robert~M Owen, Emily~C Dykhuizen, and Laura~L
  Kiessling.
\newblock Selective tumor cell targeting using low-affinity, multivalent
  interactions.
\newblock {\em ACS chemical biology}, 2(2):119--127, 2007.

\bibitem{kiessling2}
Christopher~W Cairo, Jason~E Gestwicki, Motomu Kanai, and Laura~L Kiessling.
\newblock Control of multivalent interactions by binding epitope density.
\newblock {\em Journal of the American Chemical Society}, 124(8):1615--1619,
  2002.

\bibitem{francisco-pnas}
F.~J. Martinez-Veracoechea and D.~Frenkel.
\newblock Designing super selectivity in multivalent nano-particle binding.
\newblock {\em Proceedings of the National Academy of Sciences},
  108(27):10963--10968, 2011.

\bibitem{treloar1946statistical}
Leslie Ronald~George Treloar.
\newblock The statistical length of long-chain molecules.
\newblock {\em Rubber Chemistry and Technology}, 19(4):1002--1008, 1946.

\bibitem{wick2000self}
Collin~D Wick and J~Ilja Siepmann.
\newblock Self-adapting fixed-end-point configurational-bias monte carlo method
  for the regrowth of interior segments of chain molecules with strong
  intramolecular interactions.
\newblock {\em Macromolecules}, 33(19):7207--7218, 2000.

\bibitem{mooij1994overlapping}
GCAM Mooij and D~Frenkel.
\newblock The overlapping distribution method to compute chemical potentials of
  chain molecules.
\newblock {\em Journal of Physics: Condensed Matter}, 6(21):3879, 1994.

\bibitem{siepmann1992configurational}
J{\"o}rn~Ilja Siepmann and Daan Frenkel.
\newblock Configurational bias monte carlo: a new sampling scheme for flexible
  chains.
\newblock {\em Molecular Physics}, 75(1):59--70, 1992.

\end{thebibliography}

\end{document}